%% file: main.tex
\documentclass[11pt]{article}
\pdfoutput=1

\usepackage[preprint]{acl}

\usepackage{comment}
\usepackage{times}
\usepackage{latexsym}
\usepackage{booktabs}
\usepackage{amsfonts}
\usepackage{caption}
\usepackage{subcaption}
\input{math_commands.tex}

\input{header.tex}

\usepackage[T1]{fontenc}

\usepackage[utf8]{inputenc}

\usepackage{microtype}

\usepackage{inconsolata}

\usepackage{graphicx}

\usepackage{multirow}

\title{When Benchmarks Talk: Re-Evaluating Code LLMs with\\Interactive Feedback}

\newcommand*\samethanks[1][\value{footnote}]{\footnotemark[#1]}

\author{Jane Pan\textsuperscript{1}\thanks{Equal Contribution.} \quad Ryan Shar\textsuperscript{2}\samethanks \quad Jacob Pfau\textsuperscript{1} \quad Ameet Talwalkar \textsuperscript{2} \quad He He \textsuperscript{1}\thanks{Co-senior Authors.} \quad Valerie Chen \textsuperscript{2}\samethanks\\
\textsuperscript{1}New York University
\textsuperscript{2}Carnegie Mellon University\\
\texttt{\href{mailto:jane.pan@nyu.edu}{jane.pan@nyu.edu}}
}

\begin{document}
\maketitle

\begin{abstract}
\input{sections/0_abstract}

\end{abstract}


\input{figtext/teaser}

\input{sections/1_introduction}

\input{sections/3_experiments}

\input{sections/4.1_rankings}

\input{sections/4.2_feedback_quality}

\input{sections/4.3_steerability}
\input{sections/5_related_work}

\input{sections/7_conclusion}
\input{sections/6_discussion}
\bibliography{rebib_custom}
\newpage
\appendix
\input{sections/appendix}

\end{document}

%% file: math_commands.tex
\usepackage{amsmath,amsfonts,bm}
\usepackage{graphicx}

\def\eqref#1{equation~\ref{#1}}

\def\1{\bm{1}}

\DeclareMathAlphabet{\mathsfit}{\encodingdefault}{\sfdefault}{m}{sl}
\SetMathAlphabet{\mathsfit}{bold}{\encodingdefault}{\sfdefault}{bx}{n}



%% file: header.tex
\providecommand{\bluetext}[1]{
    {\protect\color{blue}{#1}}
}
\providecommand{\orangetext}[1]{
    {\protect\color{orange}{#1}}
}
\providecommand{\tantext}[1]{
    {\protect\color{red}{#1}}
}

\newcommand{\vanilla}{\textsc{Static}}
\newcommand{\para}{\textsc{Paragraph}}
\newcommand{\sent}{\textsc{Sentence}}
\newcommand{\ir}{\textsc{Query Rephrasing}}
\newcommand{\cf}{\textsc{Code Feedback}}
\newcommand{\baseline}{\textsc{Self-Critique Baseline}}

\newcommand{\gpt}{\texttt{GPT-4o}}
\newcommand{\sonnet}{\texttt{Sonnet-3.5}}
\newcommand{\deepseek}{\texttt{Deepseek-V3}}
\newcommand{\aya}{\texttt{Aya}}
\newcommand{\reka}{\texttt{Reka}}
\newcommand{\gemmaSmall}{\texttt{Gemma-7B-it}}
\newcommand{\gemmaLarge}{\texttt{Gemma-2-27B-it}}
\newcommand{\llama}{\texttt{Llama-3.1-8B-Instruct}}
\newcommand{\qwenSmall}{\texttt{Qwen2.5-Coder-7B-Instruct}}
\newcommand{\qwenLarge}{\texttt{Qwen2.5-Coder-32B-Instruct}}

\newcommand{\user}{user}
\newcommand{\cm}{code model}

\newcommand{\nmodels}{$10$}
\newcommand{\lqf}{directionally incorrect}
\newcommand{\hqf}{directionally correct}

%% file: sections/0_abstract.tex
Programming is a fundamentally interactive process, yet coding assistants are often evaluated using static benchmarks that fail to measure how well models collaborate with users. 
We introduce an interactive evaluation pipeline to examine how LLMs incorporate different types of feedback in a collaborative setting. Specifically, we perturb static coding benchmarks so that the code model must interact with a simulated user to retrieve key information about the problem.
We find that interaction significantly affects model performance, as the relative rankings of $10$ models across $3$ datasets often vary between static and interactive settings, despite models being fairly robust to feedback that contains errors.
We also observe that even when different feedback types are equally effective with respect to performance, they can impact model behaviors such as (1) how models respond to higher- vs. lower-quality feedback and (2) whether models prioritize aesthetic vs. functional edits. 
Our work aims to ``re-evaluate'' model coding capabilities through an interactive lens toward bridging the gap between existing evaluations and real-world usage.

%% file: figtext/teaser.tex
\begin{figure}[!t]
    \centering
    \includegraphics[width=0.98\columnwidth]{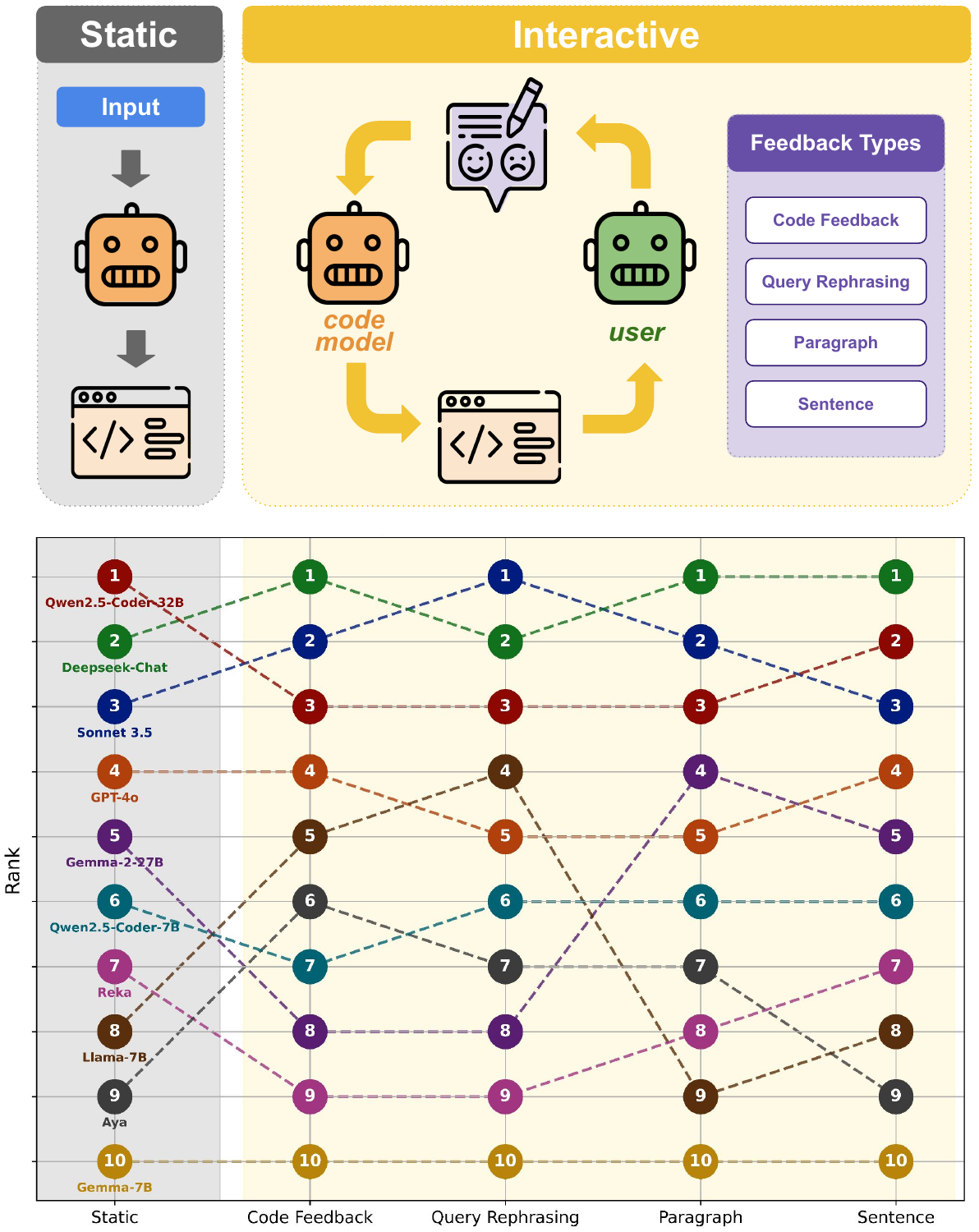}
    \caption{While most existing benchmarks statically evaluate LLM coding capabilities, code LLMs are used interactively in practice. We introduce an evaluation pipeline that evaluates \cm s in an interactive setting (top). Across three datasets, such as LiveCodeBench (bottom), we find that interactively evaluating models with different feedback types (\cf, \ir, \para, and \sent) leads to different rankings when compared to static evaluation.
    }
    \vspace{-5pt}
    \label{fig:teaser}
\end{figure}

%% file: sections/1_introduction.tex
\section{Introduction}

\input{figtext/main_figure}

Programming with a language model is a highly collaborative process, where developers interact with \cm s to provide updated information about initially underspecified requests or critique the output of the \cm.
Thus, giving and receiving feedback are critical elements of the process in which programmers use \cm{}s~\cite{chidambaram2024socratic}. 
For example, chat interfaces like ChatGPT~\citep{chatgpt} or the chat panel of Github Copilot~\cite{copilot} facilitate multi-turn conversations in which programmers can iteratively refine a piece of code by providing additional context and details to the LLM ~\citep{kalla2023study,xiao2023devgpt}.

Despite the popularity of these tools, existing static benchmarks that measure task performance often rely on a simple input-output configuration, where the question is well defined and the model is asked to generate the whole completion in one shot~\citep{chen2021evaluating,austin2021program,jain2024livecodebench, white2024livebench}.
While this relatively simplistic setting is scalable and enables efficient evaluation, it does not capture how developers realistically use models to write code~\citep{mozannar2024realhumaneval}. 
While recent benchmarks have begun to explore how interactive settings can lead to performance gains in coding applications~\cite{wang2023mint}, they assume a single form of natural language feedback. 
In practice, developers provide many forms of feedback when implementing code~\citep{chidambaram2024socratic}, which can range from binary feedback on the correctness of the code to suggesting direct changes to the code.

Building evaluations of programming assistants that more closely mimic this setting enables a better understanding of model behavior and potential pitfalls in the interactive setting, such as model capability with respect to processing feedback, the effects of different feedback types, or model robustness. 
We bridge the gap between existing evaluations and real-world use cases by benchmarking how different feedback types impact model behavior in a simulated interactive setting (Figure~\ref{fig:teaser}, top). 
We propose an evaluation method that transforms a static coding benchmark into an interactive, collaborative one (Figure~\ref{fig:main_figure}). 
The pipeline components include input obfuscation to create underspecified problems to induce collaboration, a simulated user for scalability, and multiple types of feedback (\cf, \ir, \para, and \sent).

Across \nmodels{} coding models ($6$ open and $4$ closed) and $3$ coding benchmarks, we find that relative performance between models often changes between static and interactive settings (Figure~\ref{fig:teaser}, bottom), suggesting that models perform differently in the static vs. interactive settings.
Beyond performance-based metrics, we analyze important components of model-user interactions, including feedback quality and code model steerability.
We use these insights to investigate the effects of different feedback types.
For example, we find that \para{} and \cf{} tend to lead to the highest performance boost compared to other feedback types (e.g., \sent{} or \ir). 
However, when considering the effect of feedback quality on performance, we find that unlike \para, \cf 's higher-quality feedback makes output worse more frequently than its lower-quality counterpart for stronger models.
Furthermore, \para{} leads to more surface-level edits than \cf, whereas users may prefer variation in model behavior to be robust to changes in feedback type.

Our work provides a new approach to investigating the downstream ramifications of different interactive programming settings and their effects on model behavior.
We open-source our evaluation pipeline\footnote{Our code is publicly available at \url{https://github.com/janepan9917/WhenBenchmarksTalk/}.}, which makes it easy to add static benchmarks and turn them into interactive ones, to facilitate the evaluation of more models and datasets.

%% file: figtext/main_figure.tex
\begin{figure*}[ht]
    \centering
    \includegraphics[width=0.98\textwidth]{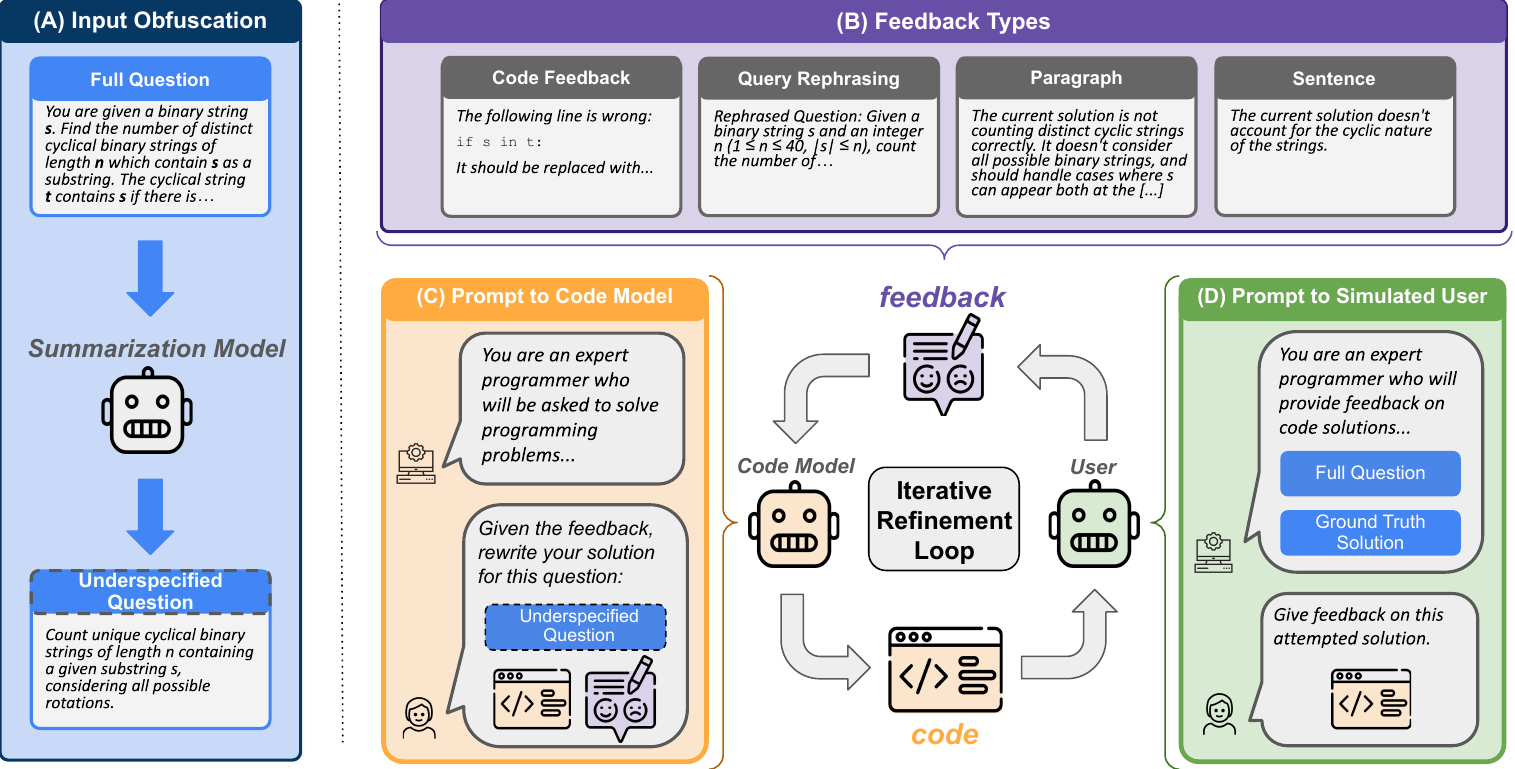}
    \caption{Overview of our interactive pipeline for coding evaluation. 
    (A) We obfuscate the input of existing fully specified datasets to reflect how programmers tend to underspecify requests to LLMs (e.g., via docstrings or comments) in practice. (B) As developers may interact with models in a variety of ways, we explore $4$ different feedback types and introduce a pipeline that mimics the iterative refinement loop that programmers often use with chat models, (C) where the \cm{} generates a solution using feedback on its previous solution, (D) and the \user{} provides updated feedback to the \cm.
    }
    
    \label{fig:main_figure}
\end{figure*}

%% file: sections/3_experiments.tex
\section{Methodology}
Figure \ref{fig:main_figure} provides an overview of our pipeline for interactive evalution, which is driven by an iterative refinement loop in which the \user{} model and \cm{}  interact.
At each step of the loop, the \cm{} is given the obfuscated programming question in natural language, the code from the previous attempt, and the \user{} model's feedback on the previous attempt. Each iterative refinement loop lasts $5$ steps for each question, terminating early if a correct solution is reached in less than $5$ steps.

\subsection{Dataset Transformation}
\label{sec:dataset_transform}
Our pipeline is designed to transform static coding questions into interactive ones. 
We describe our selection criteria for the datasets used in this work, as well as the input obfuscation protocol that enables collaboration between the \user{} and \cm{}.

\paragraph{Input obfuscation.} 
To ensure that the \user{} and \cm{} collaborate, we obfuscate the input given to the \cm{} to remove critical information from the input (Figure \ref{fig:main_figure}A). 
This induces an information asymmetry, akin to that which might exist in practice, that forces the \cm{} to rely on feedback from the \user{} to recover key details about the full problem specification.
We underspecify the input to the \cm{} by using \sonnet{} to summarize the original questions, which often contain significant detail about desired behaviors and potential edge cases, into one-sentence summaries.

To illustrate input obfuscation, consider a question from APPS (Interview) which asks how long it will take for two flight attendants to serve lunch to a customer in a given seat. The original question includes details about the flight attendant's serving speed, the order they traverse the rows of the plane, and the serving order of seats in the row. 
The summarized form of this question might be \textit{``Calculate the time it takes for a passenger in a specific seat to receive their lunch on an airplane with an infinite number of rows, given the serving pattern of two flight attendants moving from front to back,''} omitting some of the key details required to fully solve the question (e.g., row and seat order).
We use this as a running example in the following section.

\paragraph{Datasets.} 
We select challenging datasets with lengthy problem descriptions and available ground-truth solutions.
We use the Interview and Introductory levels of APPS~\citep{hendrycks2021apps}, ClassEval~\citep{du2023classeval}, and the Easy, Medium, and Hard levels of LiveCodeBench~\citep{jain2024livecodebench}.
We randomly sample $200$ examples from APPS Interview, $200$ examples from APPS Introductory, and $75$ examples from ClassEval. 
We use $70$ examples from LiveCodeBench (across all three difficulty levels). 
More details on the datasets can be found in Appendix \ref{app:datasets}.

\subsection{Feedback Types} 
Existing work on developer-\cm{} interactions shows that programmer feedback is diverse~\citep{chidambaram2024socratic}. 
Following their results, we explore multiple variants of feedback types outside of generic natural language feedback and investigate four categories of feedback: \para, \sent, \ir, and \cf{} (Figure \ref{fig:main_figure}B).

The \sent{} and \para{} feedback types are ones where the \user{} model provides feedback using natural language.
These feedback styles mimic the inputs used in chat-based interfaces, where users respond to the model via a chat window. In our setting, the \user{} is prompted to only use a sentence or paragraph for their response. 
An example of \sent{} feedback for the airplane question might be \textit{``The current solution doesn't follow the seat serving order f-e-d-a-b-c,''} while \para{} feedback tends to have specific critiques of the algorithm with sentences like \textit{``It doesn't account for the two flight attendants serving simultaneously. It should first calculate the number of complete 4-row blocks served, then handle the remainder.''} 

\ir{} and \cf{} feedback aim to replicate common feedback styles from developers~\citep{chidambaram2024socratic}. 
In \ir{}, the \user{} is prompted to rewrite a similar-length version of the underspecified question with additional details required for the \cm{} to find a solution. 
We constrain the length to mimic how real users rephrase their inputs when providing feedback and to prevent the \user{} from simply copying in the full question.
An example of \ir{} in the airplane question might be \textit{``Question: Calculate the time for a passenger to receive lunch on a plane where two flight attendants serve food. Attendants start at rows 1 and 3, move forward by 2 rows after serving. They serve right side (f to d) then left side (c to a) of each row. Output the waiting time in seconds.''}
\cf{} prompts the \user{} to directly indicate which lines of code are incorrect and suggest alternate code snippets. 
An example of \cf{} in the airplane question might be \textit{``The function get\_time\_to\_lunch(seat, num\_attendants) should be get\_time\_to\_lunch(seat) as the number of attendants is always 2.''}
Appendix \ref{app:feedback} provides additional examples of each type of feedback.

\subsection{Code Models}
We select a total of \nmodels{} \cm s, spanning both open-source and closed models and a wide range of capabilities and parameter sizes. We selected the following open-source models for their ability to follow user instructions, their range of parameter sizes, and overall coding capability:  \deepseek~\citep{deepseekai2024deepseekv3technicalreport}, \gemmaSmall~\citep{gemmateam2024gemmaopenmodelsbased}, \gemmaLarge{}~\citep{gemmateam2024gemma2improvingopen}, \llama{}~\citep{grattafiori2024llama3herdmodels}, \qwenSmall{}~\citep{hui2024qwen25codertechnicalreport}, and \qwenLarge{}~\citep{hui2024qwen25codertechnicalreport}. 
We select the following closed models for their commercial adoption and performance on existing static benchmarks:
\aya{}~\citep{aya2024modelinstructionfinetuned},, \gpt{}~\citep{openai2024gpt4ocard}, \reka{}~\citep{rekateam2024rekacoreflashedge}, and \sonnet{}~\citep{claude}. The parameters used to query each model are provided in Appendix~\ref{app:models}.

\paragraph{Code model prompts.} 
We prompt the \cm s with their previous solution and the \user{} feedback on that solution. 
We do not provide a history of all \cm{} interactions with the \user{}, only the most recent code attempt and \user{} feedback on the most recent attempt.
Specific instructions are given for each dataset for varying input and output formats.
All prompts are in Appendix \ref{app:prompts}.

\subsection{User Models}

Following prior work~\citep{dubois2023alpacafarm,zheng2023judging,mozannar2023simulating}, we use LLMs to scalably simulate feedback given by \user s when interacting with \cm s. 
To help close the capability gap between real-world expert users and LLMs, we give the \user{} model access to the original fully-specified question, as well as a ground-truth solution. 
This allows the simulated \user{} to produce higher-quality feedback more often. 
The \user{} model prompt also includes instructions to avoid leaking the exact solution in its responses to the \cm.
We only constrain the formatting style of the feedback, allowing the \user{} to choose the content of criticism (e.g. input/output formatting, algorithmic correctness, code style).
We choose \sonnet{} due to its high performance on static coding benchmarks and strong reasoning capabilities, but verify that other \user{} models (e.g., \texttt{GPT-4o-mini}) are also able to improve performance over \baseline{} (see Section \ref{sec:static_vs_interactive_perf} for details).
These results, along with prompts for the \user{} can be found in Appendix \ref{app:prompts}.

%% file: sections/4.1_rankings.tex
\section{Static vs. Interactive Performance}
\label{sec:static_vs_interactive_perf}

\input{tables/perf_per_dataset_per_feedback}

\input{figtext/ranking_changes}

We compare the performance of models in static and interactive evaluation settings.
To measure the effectiveness of different feedback types, we compare against the \vanilla{} and the \baseline{} settings.
The \vanilla{} setting of each dataset evaluates the \cm{} on the original, fully specified questions.
In this setting, the \cm{} is not given any feedback and the first attempt is used as the final output.
To match the test-time compute of the interactive settings, the \baseline{} setting uses five iterations of self-critique to generate feedback using the underspecified question and the output from the previous step ~\citep{madaan2023selfrefineiterativerefinementselffeedback}.
For this setting, no additional information of the original question or solution is given and no \user{} is involved.

\subsection{Performance Metrics}
We use \textbf{test case accuracy (TCA)} to evaluate model performance. In ClassEval, we combine the set of function tests and class tests to measure TCA. To measure the distances between two rankings $\sigma_A$ and $\sigma_B$ of length $n$, we use a normalized variant of \textbf{Spearman's Footrule } ($\tilde{F}$ $:\ \to [0, 1]$):
\begin{align*}
    \tilde{F}(\sigma_A, \sigma_B) &= \frac{ \sum^n_{i=1} |\sigma_A(i) - \sigma_B(i)|}{\max_{\sigma, \sigma'}  \sum^n_{i=1} |\sigma(i) - \sigma'(i)|}
\end{align*}
where $\sigma, \sigma'$ are any ranking of length $n$.
For a perfectly correlated pair of rankings, $\tilde{F}=0$; for uncorrelated rankings, $\tilde{F}=0.73$; for perfectly anti-correlated rankings, $\tilde{F}=1$.
Appendix \ref{app:performance_metrics} contains details on how we derive these metrics and thresholds for correlation.

\subsection{Results}

\paragraph{Feedback can recover performance comparable to or even exceeding the \vanilla{} setting.}
Table \ref{tables:perf_per_dataset_setting} shows the performances averaged across all models for each dataset and feedback type.
Comparing \vanilla{} to \baseline{} shows that our input perturbation often obfuscates the problem, as \baseline{}  usually underperforms the \vanilla{} setting, and feedback is often required to achieve performance comparable with the \vanilla{} setting (as with LiveCodeBench).
Moreover, interacting with feedback may also allow \cm s to surpass \vanilla{} performance (as with APPS and ClassEval). 
This may be because the \user{} may supply not only additional specifications about the problem, but also guidance with respect to general problem-solving or programming capabilities.

\paragraph{\cf{} and \para{} improve performance the most.}
\cf{} and/or \para{} are consistently the most effective at improving model performance in the underspecified setting (Table \ref{tables:perf_per_dataset_setting}). 
Compared to others, these feedback types tend to be longer and thus may encapsulate more helpful information to the \cm.
The weaker performing feedback types are \sent{} and \ir, the latter of which struggles the most on LiveCodeBench.
This may stem from the fact that LiveCodeBench is not in the training sets of most models (due to its problem cut-off date); for popular datasets,  the \cm{} may sometimes recognize the full question using the \user 's \ir{}, leading to improved performance.

\input{figtext/effect_of_feedback_quality}

\paragraph{Models perform differently in static vs. interactive settings.} 
Figure \ref{fig:ranking_changes} plots the relative rankings (measured by TCA) across static and interactive settings. 
While we generally observe permutations in rankings when comparing \vanilla{} and interactive, LiveCodeBench demonstrates the most variance, with some models changing $4$ ranks between \vanilla{} and interactive settings.

To understand how rank changes vary between static vs. interactive settings, we calculate the normalized Spearman's Footrule distances between the \vanilla{} and interactive settings (Table \ref{tables:normalized-spearmans-footrule-dist}).
All three datasets demonstrate relatively weak positive correlation in many interactive settings; for instance, for \cf, $\tilde{F}$ ranges from 0.222 to 0.346 across the three datasets.
Generally, top models tend to be consistently high across feedback types, whereas weaker models tend to demonstrate more variance in rank.

\input{tables/feedback_quality_max_step}

%% file: tables/perf_per_dataset_per_feedback.tex
\begin{table*}[t]
\centering
\resizebox{\textwidth}{!}{
\begin{tabular}{ccccccc}
\toprule 
\textbf{Dataset} & \textbf{\vanilla} & \textbf{\baseline} & \textbf{\para} & \textbf{\sent} & \textbf{\cf} & \textbf{\ir} \\ \hline
APPS & 0.335 (0.003) & 0.034 (0.001) & 0.381 (0.004) & 0.271 (0.003) & 0.428 (0.004) & 0.289 (0.003) \\
LCB & 0.699 (0.008) & 0.274 (0.008) & 0.655 (0.009) & 0.611 (0.009) & 0.631 (0.009) & 0.183 (0.008) \\
ClassEval &  0.714 (0.002) &  0.483 (0.006) & 0.679 (0.006) & 0.642 (0.006)& 0.759 (0.006) & 0.693 (0.005)\\
 \bottomrule
\end{tabular}
}
\caption{
    Test case accuracy and standard error of each setting in APPS, LiveCodeBench (LCB), and ClassEval, averaged across all \cm s. We find that feedback can recover performance comparable to or even exceeding the \vanilla{} setting.
    }
    \label{tables:perf_per_dataset_setting}
\end{table*}


%% file: figtext/ranking_changes.tex
\begin{figure*}[t]
    \centering
    \includegraphics[width=0.98\textwidth]{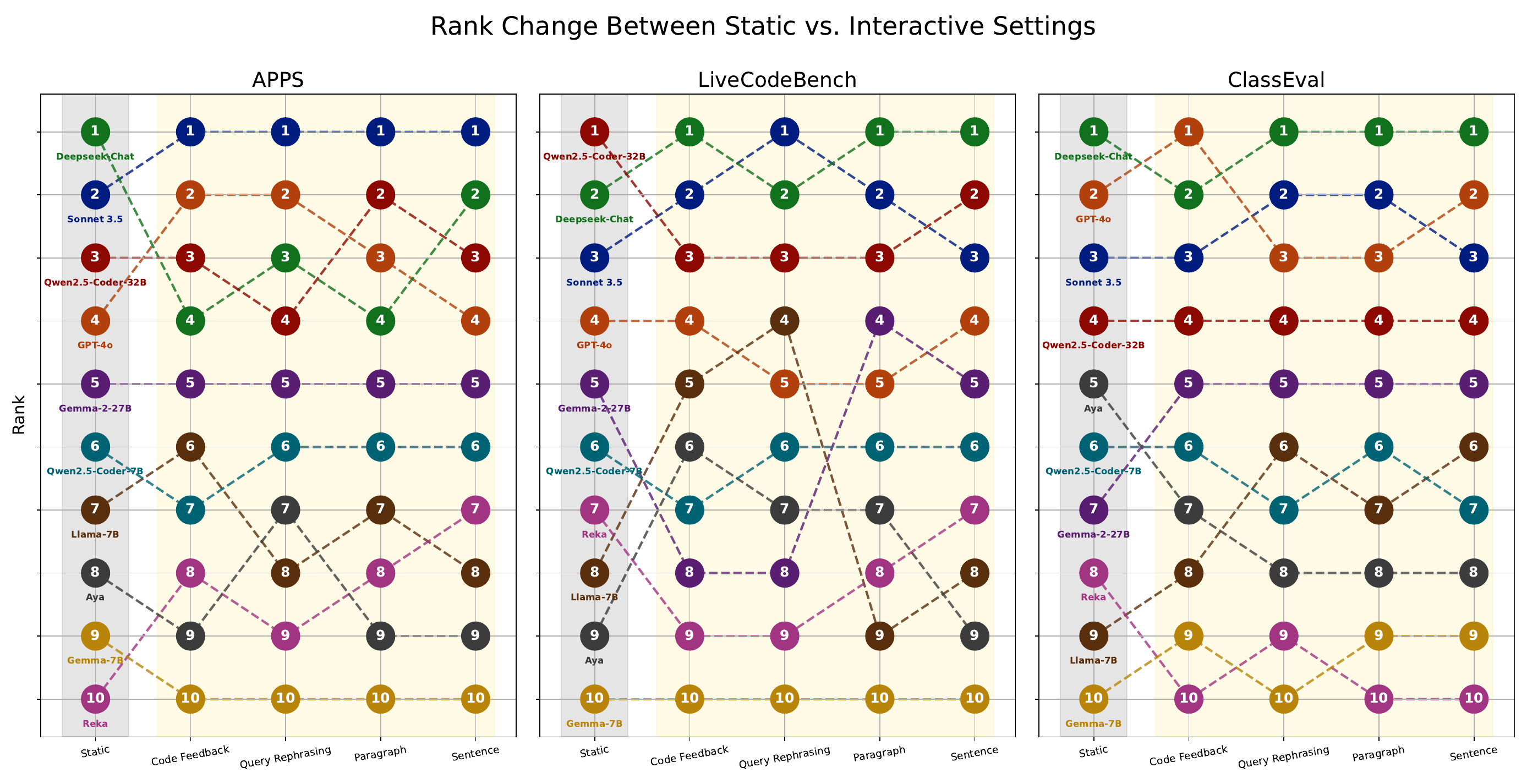}
    \caption{Rank changes between static and interactive settings across $3$ datasets--- APPS, LiveCodeBench, and ClassEval. We stratify interactive settings by feedback type (\cf, \ir, \para, and \sent), and observe changes in rankings across all datasets and interactive settings.
    }
    
    \label{fig:ranking_changes}
\end{figure*}

%% file: figtext/effect_of_feedback_quality.tex
\begin{figure*}[ht]
    \centering
    \includegraphics[width=0.98\textwidth]{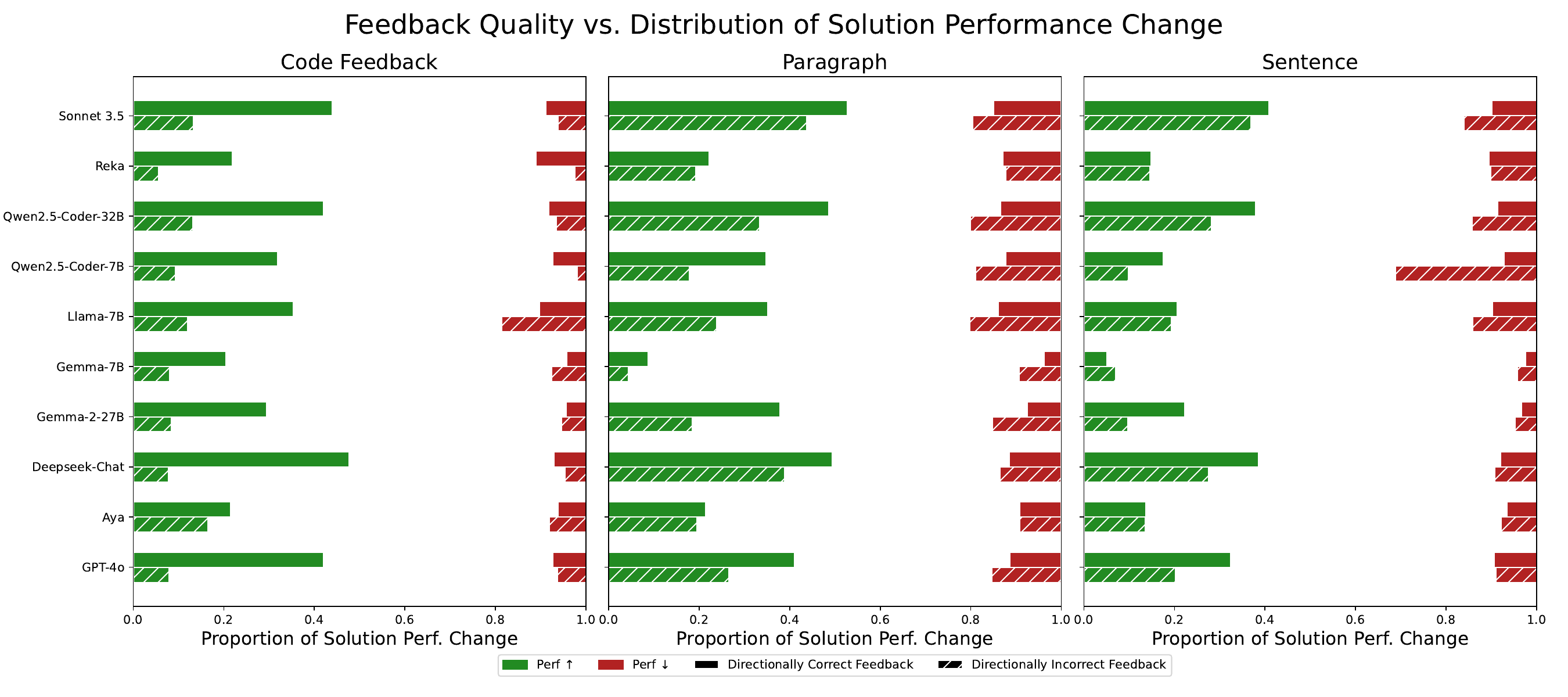}
    \caption{Distribution of performance change across feedback types and directional correctness. We split solutions into post-feedback performance gains (green) or losses (red) and observe that models can still benefit from \lqf{} feedback, and that \hqf{} \cf{} sometimes increases the rate of post-feedback performance loss. 
    }
    \label{fig:effect_of_fq}
\end{figure*}

%% file: tables/feedback_quality_max_step.tex
\begin{table}[t]
\centering
\resizebox{\columnwidth}{!}{
\begin{tabular}{cccc}
\toprule 
\textbf{Dataset} & \para & \sent & \cf \\ \hline
APPS &  0.91 (0.01) & 0.89 (0.01) & 0.94 (0.01)\\
 \hline 
LCB & 0.92 (0.01)  & 0.88 (0.02)  &  0.79 (0.03)\\
 \hline
ClassEval &  0.96 (0.01) & 0.91 (0.01) & 0.77  (0.02)\\
 \bottomrule
\end{tabular}
}
\caption{
    Average rate of directional correctness with standard error for each dataset and setting.
    }
    \label{tab:feedback_quality_max_step_number}
\end{table}

%% file: sections/4.2_feedback_quality.tex
\section{Feedback Quality}
To understand whether the ranking changes in Section \ref{sec:static_vs_interactive_perf} are due to variations in feedback quality across models, we develop a proxy for feedback quality and examine its effect on how \cm s interact with feedback.

\subsection{Quality Metrics}
Previous works infer feedback quality by the feedback's effect on performance \cite{zhang2023clarifynecessaryresolvingambiguity}.  
Instead of relying only on performance, we classify feedback by \textbf{directional correctness}, a binary value of whether it accurately claims that the code solution was correct or incorrect. 
For instance, if the solution is incorrect (i.e. has $\textsc{TCA} < 1$), but the feedback claims that the solution is correct, then we consider the feedback \lqf.
However, if the solution is correct (i.e. has $\textsc{TCA} = 1$) and the feedback claims that it is correct, we consider it \hqf.

We automate the classification via \gpt{} on \sent, \para, and \cf{}.\footnote{\ir{} does not provide direct feedback on the solution, so it is not eligible for our quality metric.}
Appendix \ref{app:fq} discusses other feedback quality metrics we considered, as well as additional information on the classification protocol. 

\subsection{Results}

\paragraph{Directional correctness is consistently high across models and feedback types.}
Table \ref{tab:feedback_quality_max_step_number} compares average directional correctness by feedback type, which does not vary greatly across models or feedback types and often reaches above $0.8$.
This suggests that the feedback is high enough quality to compare across models and feedback types.
Although \para{} and \cf{} feedback induce the highest performances, \para{} feedback tends to have the highest directional correctness, whereas \cf{} tends to have the lowest.

\paragraph{Code models are generally robust to \lqf{} feedback.}
Figure~\ref{fig:effect_of_fq} shows the distribution of solutions whose performances improve versus decrease when comparing \hqf{} feedback to \lqf{} feedback. 
Although \hqf{} feedback has a higher rate of solutions whose performances improve, the rate of directionally \textit{incorrect} feedback that results in improved performance is still substantial.
For instance, \sonnet{} has equal rates of improved performance after either \hqf{} or \lqf{} \para{} and \sent{} feedback.

\paragraph{For stronger models, \hqf{} \cf{} tend to worsen post-feedback solutions than \lqf{} \cf.} While all \hqf{} feedback have roughly similar effects on the rate of improved post-feedback solutions, \para{} and \sent{} also decrease the proportion of worse post-feedback solutions (Figure \ref{fig:effect_of_fq}, center and right).
\cf{} is the only feedback type where stronger models (e.g., \sonnet, \gpt, \qwenLarge) are more likely to generate a worse solution when given \hqf{} feedback rather than \lqf{} (Figure \ref{fig:effect_of_fq}, left).

%% file: sections/4.3_steerability.tex
\input{figtext/feedback_steerability_barchart}

\section{Model Steerability}

We extend our analysis beyond performance to investigate \textbf{steerability}, or how much a \cm{} adjusts its previous solution in response to feedback.
We show that drops in performance in interactive settings are likely due to ineffectively incorporating feedback, rather than outright ignoring feedback in the next iteration of the solution.

\input{figtext/edit_distance_vs_steerability_heatmap}

\subsection{Steerability Metrics}
We evaluate model steerability on APPS and LiveCodeBench\footnote{We do not evaluate on ClassEval as their evaluation utilities do not report exactly which test cases pass or fail.} and consider two metrics of change between solution iterations. 
Firstly, we use \textbf{Levenshtein edit distance} to evaluate surface-level changes between consecutive versions of code. 
Secondly, we count the number of \textbf{changes in test-case behavior} --- with respect to whether incorrect test cases flip to correct or vice-versa --- to evaluate behavioral-level adjustments to the code.
We refer to the former as \textit{surface-level steerability} and the latter as \textit{behavioral steerability}.

\subsection{Results}
\paragraph{\para{} feedback is associated with higher behavioral-level and surface-level steerability across all models.} 
Figure \ref{fig:feedback_steerability_barchart} plots each feedback type by the behavioral (top) or surface-level (bottom) steerability it induces.
\para{} and \cf{} score the highest in behavioral steerability (changing 21.8\% and 19.3\% of test cases on average), while \para{} score the highest in surface-level steerability (with an average edit distance of 445.6 characters).
Notably, \para{} is also one of the highest-performing feedback styles, suggesting that it induces effective changes in the code solution on both the behavioral and surface levels.

\paragraph{Weaker models tend to make surface-level rather than effective behavioral-level changes.} 
Figure \ref{fig:edit_distance_vs_steerability_heatmap} shows how \cm s change their previous solutions on APPS Interview across feedback types.
Weaker models (e.g. \gemmaSmall{} and \llama) tend to make many surface-level changes that do not greatly change the behavior of the code.
However, stronger models (e.g. \gpt{}, \sonnet, and \qwenLarge) may make relatively small edits that highly affect code behavior.

%% file: figtext/feedback_steerability_barchart.tex
\begin{figure}[t]
    \centering
    \includegraphics[width=0.98\columnwidth]{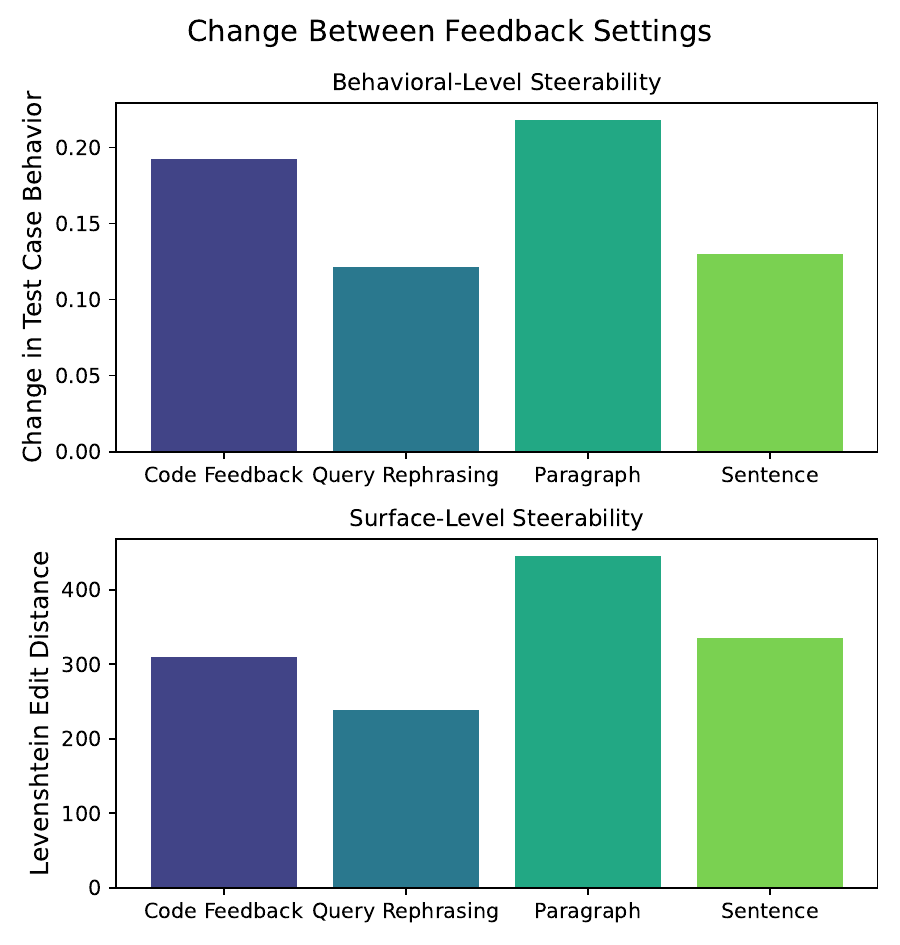}
    \caption{
         Behavioral-level (top) and surface-level (bottom) steerability by feedback type, averaged across all models for APPS and LiveCodeBench.
         \para{} feedback induces the most changes at both levels, while \cf{} leads to more behavioral changes with less aesthetic changes.
    }
    
    \label{fig:feedback_steerability_barchart}
\end{figure}

%% file: figtext/edit_distance_vs_steerability_heatmap.tex
\begin{figure*}[ht]
    \centering
    \includegraphics[width=0.98\textwidth]{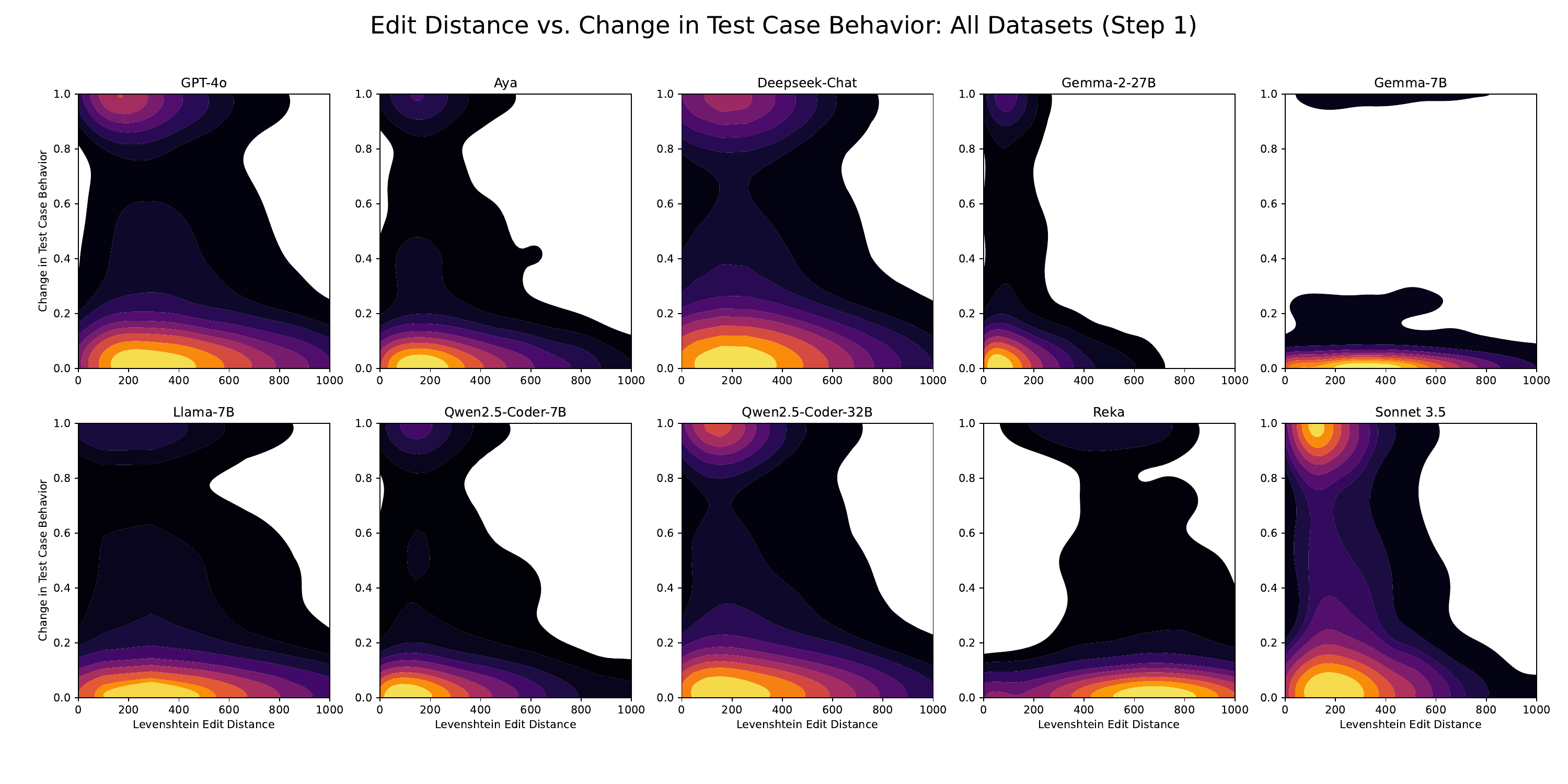}

    \caption{
        Distribution of surface-level steerability ($x$-axis) vs. behavioral steerability ($y$-axis) for all models during the first step of iterative refinement. While some models make only surface-level changes that do not induce much behavioral change in code (e.g. \gemmaSmall), others are also able to make highly effective edits that induce large changes in the behavior of the solution (e.g., \sonnet, \qwenLarge, \gpt).
    }
    \vspace{-5pt}
    \label{fig:edit_distance_vs_steerability_heatmap}
\end{figure*}

%% file: sections/5_related_work.tex
\section{Related Work}

\paragraph{Code benchmarks.}
Static benchmarks, e.g., HumanEval~\citep{chen2021evaluating} and MBPP~\citep{austin2021program}, largely focusing on interview-style programming problems, have been the most commonly used to evaluate coding capabilities~\citep{lu2021codexglue, nijkamp2022codegen,zhu2022xlcost, wang2022recode, liu2023your, jimenez2023swe, khan2023xcodeeval,yan2023codescope, cassano2023multipl, muennighoff2023octopack, dinh2023large,yang2023intercode, du2023classeval}. 
Recent live benchmarks aim to reduce contamination risks~\citep{jain2024livecodebench,white2024livebench}.
Our evaluation pipeline can convert many of these static benchmarks into an interactive one, evaluating model abilities to incorporate different types of feedback; we demonstrate this with $3$ datasets.

\paragraph{Interactive evaluation.}
As programmers are increasingly writing code collaboratively with AI chat assistants like ChatGPT~\cite{chatgpt} or Claude~\cite{claude}, many user studies have evaluated how programmers use chat assistants to write code~\citep{ross2023programmer, chopra2023conversational, kazemitabaar2023studying,xiao2023devgpt,nam2024using,mozannar2024realhumaneval,chidambaram2024socratic}, typically employing only a few models in the study ($\leq 3$).
While new platforms evaluate model coding capabilities at scale by collecting human preferences~\citep{chiang2024chatbot, chi2024copilot}, it remains challenging to understand the fine-grained effects of feedback.
We create a benchmark with \emph{simulated} users to enable scalable evaluation of the nuances of feedback in interactive coding settings, while drawing from insights of existing \emph{human} studies (e.g., common types of feedback~\citep{chidambaram2024socratic}  and tendencies to underspecify inputs~\citep{xiao2023devgpt}).

Prior work has explored interactive benchmarks with simulated users for various applications, such as tool use \citep{yao2024taubenchbenchmarktoolagentuserinteraction}, creative tasks \citep{jia2024simulbenchevaluatinglanguagemodels},  coding \citep{wang2024mintevaluatingllmsmultiturn, shao2025collaborativegymframeworkenabling}, and other collaborative contexts \citep{wu2023autogenenablingnextgenllm}. 
Our benchmark extends them by introducing diverse forms of \user{} feedback and investigating their effect on feedback quality and model steerability. 
We also enforce collaboration between the simulated \user{} and \cm{} via input obfuscation, aligning with real-world use cases where the user's input to the model may be underspecified.

%% file: sections/7_conclusion.tex
\section{Conclusion}
We propose a new approach to evaluating \cm s by introducing an interactive pipeline where the \cm{} must collaborate with a simulated \user{} to solve underspecified coding problems with different feedback types.
We find that the relative performances of models change radically between static and interactive settings.
We analyze key elements of model-user interactions, such as feedback quality and model steerability, to provide insights into the downstream effects of feedback type on model behavior and feedback effectiveness.
Our work bridges the gap between existing static benchmarks and real-world usage, and we hope to inspire future work on scalable methods for evaluating models in a collaborative setting.

%% file: sections/6_discussion.tex
\section{Limitations}

In this work, the focus of our experiments is on three diverse code benchmarks to demonstrate the generality of our pipeline. 
However, given the expansive set of static benchmarks, our results may not encompass the full set of observations one might obtain from considering more varied datasets (e.g., non-Python coding questions).
On the evaluation front, since we do not explicitly compare LLM responses to human-generated feedback on the extensive set of modified questions, we focus on trends of model behavior as a response to different feedback types, rather than specific degrees of change.

Additionally, the feedback types studied in this work are not fully comprehensive, and we do not claim that the user model's feedback is necessarily representative of actual human users.
Rather, we use a simulated \user{} to scalably examine how LLMs react to feedback in a collaborative setting, not to realistically imitate how expert humans use LLMs to program.
For instance, users may mix feedback types within a single interaction, whereas we fix the feedback type across rounds of iterative refinement to isolate the effects of an individual feedback type.
Recent works also study whether LLMs can proactively seek user feedback via clarification questions or other interactions \cite{zhang2023clarifynecessaryresolvingambiguity,zhang2024modelingfutureconversationturns}, whereas we only consider settings where the user initiates the feedback-giving process.
This work was not intended to exhaustively test feedback types but to highlight a new approach to understanding the downstream effects of interactive programming settings.

\subsection{Potential Risks}
Our pipeline is intended to be used as a testbed to examine model behavior in response to feedback, rather than to realistically mimic actual human usage with models. 
As such, it should not be used as an approximation of actual human users.
Other risks include overexposure to certain programming languages and natural languages, as we only include Python programming questions and English feedback.

\section{Acknowledgements}
We thank the members of the NYU ML\textsuperscript{2} group for their valuable advice, thoughts, and discussions. We also thank Vishakh Padmakumar, Austin Wang, Jens Tuyls, and Naman Jain for their feedback and comments. This project has benefited from financial support from Eric and Wendy Schmidt (made by recommendation of the Schmidt Futures program) and Open Philanthropy, and from in-kind support by the NYU High-Performance Computing Center. This material is also partially supported by the National Science Foundation under Award IIS-2340345. Any opinions, findings, and conclusions or recommendations expressed in this material are those of the author(s) and do not necessarily reflect the views of the National Science Foundation.



%% file: sections/appendix.tex
\newpage
\section{Appendix}
\label{sec:appendix}

\subsection{Additional Details on Datasets}
\label{app:datasets}
Our pipeline is designed to accommodate generic static benchmarks with some modifications. 
For instance, because ClassEval requires the model to fill in the skeleton code of a class (rather than providing explicit programming questions), we underspecify its problems by summarizing the docstrings for each method.
Likewise, LiveCodeBench does not provide ground-truth solutions, so we generate solutions for LiveCodeBench by sampling twice from \sonnet{}; if a correct solution is generated, we use it as the ground-truth solution for the question. 
(If not, we do not use the question as our pipeline requires the presence of a ground-truth solution in the \user{} prompt.)  

For evaluation of APPS and LiveCodeBench, Table \ref{tables:perf_per_dataset_setting}, Table \ref{tab:feedback_quality_max_step_number}, and Figure \ref{fig:ranking_changes} average across difficulty levels for brevity.

\subsection{Example Feedback}
\label{app:feedback}
\input{tables/feedback_examples}

We provide sample feedback from \sonnet{} in response to a proposed solution. All feedback types are in response to the same question (Table \ref{tables:sample-feedback}).

\subsection{Additional Details on Models}\label{app:models}

We obtained the weights for google/gemma-7b-it (\gemmaSmall{}) from Hugging Face at \url{https://huggingface.co/google/gemma-7b-it}, meta-llama/Meta-Llama-3.1-8B-Instruct (\llama{}) from huggingface at \url{https://huggingface.co/meta-llama/Llama-3.1-8B-Instruct}, and Qwen/Qwen2.5-Coder-7b-Instruct (\qwenSmall{}) from Hugging Face at \url{https://huggingface.co/Qwen/Qwen2.5-Coder-7B-Instruct}. 
We run each of the models on a single L40S GPU. We use a temperature setting of $0.9$, $4096$ max tokens, and the ``do\_sample'' setting enabled. 
We use $\langle$end\_of\_turn$\rangle$, $\langle$|eot\_id|$\rangle$, $\langle$|im\_end|$\rangle$ as the EOS token for \gemmaSmall{}, \llama{}, and \qwenSmall{} respectively.

We use Together AI (\url{https://api.together.xyz/}) to run the large open-weight models google/gemma-2-27b-it (\gemmaLarge{}) and Qwen/Qwen2.5-Coder-32B-Instruct (\qwenLarge{}). 
We set ``n\_sample'' to $1$ when generating code solutions and limit the max number of tokens to $4096$.
The weights for these models can be found at \url{https://huggingface.co/google/gemma-2-27b-it} and \url{https://huggingface.co/Qwen/Qwen2.5-Coder-32B-Instruct} respectively.

We access \texttt{c4ai-aya-expanse-32b} (\aya{}) through the Cohere API at \url{https://cohere.com/research/aya}, reka-core-20240501 (\reka{}) through the Reka API at \url{https://www.reka.ai/reka-api}, and deepseek-chat (\deepseek{}) through the Deepseek API at \url{https://api-docs.deepseek.com/}.  
We use the default API settings for inference, limiting the max number of tokens to $4096$.

\gpt{} inference is done through the OpenAI API \url{https://platform.openai.com/docs/overview} and \sonnet{} inference through the Anthropic API \url{https://www.anthropic.com/api}.
We use the default API settings for inference, limiting the max number of tokens to $4096$.

\subsection{Prompts}\label{app:prompts}
We use \sonnet{} to generate \user{} feedback for out experiments, but we show that other models can produce comparable feedback (Figure \ref{fig:4o_user_w_sonnet_apps})

We provide all of the prompts for the \user{} model and \cm{}. 
All \user{} model prompts were provided with the same system prompt with the original question and code solution (Figure \ref{prompts:user-sys-prompt}). 
The \para{} prompt (Figure \ref{prompts:user-para-prompt}) and \sent{} prompt (Figure \ref{prompts:user-sent-prompt}) are given the current \cm{} solution and generate feedback constrained by output length. 
The \cf{} prompt is given the current \cm{} solution and provides a correction to specific lines of code in the solution (Figure \ref{prompts:user-code-prompt}). 
The \ir{} feedback prompt is given the current \cm{} solution and underspecified question and generates an updated version of the question with missing details (Figure \ref{prompts:user-input-refine-prompt}).

The \cm{} prompts for APPS and LiveCodeBench are given in Figure \ref{prompts:asst-vanilla_apps_lcb} and Figure \ref{prompts:asst-nl_para_sent_apps_lcb}.
The \cm{} prompts for ClassEval is given in Figure \ref{prompts:asst-classeval_vanilla} and Figure \ref{prompts:asst-nl_para_sent_classeval}.

The 11-shot prompt used to summarize APPS and LiveCodeBench questions is given in Figure \ref{prompts:summarization_apps_lcb}.
The prompt used to summarize each docstring in ClassEval is given in Figure \ref{prompts:summarization_classeval}.

\input{figtext/4o_user_plot}
\input{prompts/user_system}
\input{prompts/user_para}
\input{prompts/user_sent}
\input{prompts/user_code}
\input{prompts/user_input_refine}
\input{prompts/asst_vanilla_apps_lcb}
\input{prompts/asst_nl_para_sent_apps}
\input{prompts/asst_vanilla_classeval}
\input{prompts/asst_nl_para_sent_classeval}
\input{prompts/summarization_apps_lcb}
\input{prompts/summarization_classeval}

\subsection{Performance Metrics}
\label{app:performance_metrics}
\paragraph{Test case accuracy.}
Test case accuracy can be defined as below:
\begin{equation*}
    \textsc{TCA} = \frac{\text{\# Test Cases Passed}}{\text{Total \# Test Cases}}
\end{equation*}
\paragraph{Normalized Spearman's Footrule distance.}

The normalized Spearman's Footrule distance is:
\begin{align*}
    \tilde{F}(\sigma_A, \sigma_B) &= \frac{ \sum^n_{i=1} |\sigma_A(i) - \sigma_B(i)|}{\max_{\sigma, \sigma'}  \sum^n_{i=1} |\sigma(i) - \sigma'(i)|}
\end{align*}
Consider two rankings $\sigma_A$ and $\sigma_B$ over items $\{1, 2, \dots, n\}$. To measure the distance between them, we use Spearman's Footrule Distance, which can be thought of as the Manhattan distance between two rankings:

\begin{equation*}
    F(\sigma_A, \sigma_B) = \sum^n_{i=1} |\sigma_A(i) - \sigma_B(i)|
\end{equation*}

We normalize $F$ by its maximum possible value, $\frac{n^2}{2}$ for even $n$, to get the \textbf{Normalized Spearman's Footrule Distance}. 
\begin{align*}
    \tilde{F}(\sigma_A, \sigma_B) &= \frac{F(\sigma_A, \sigma_B)}{\frac{n^2}{2}} \\
    &= \frac{2F(\sigma_A, \sigma_B)}{n^2}
\end{align*}
where $\tilde{F}: \to [0, 1]$. In other words, $\tilde{F}=0$ indicates $\sigma_1 = \sigma_2$, whereas $\tilde{F}=1$ indicates the maximum possible distance between $\sigma_1$ and $\sigma_2$.

Now, we would like to derive the expected $\tilde{F}$ between two rankings which are completely uncorrelated. 
Let us randomly sample $\sigma_A, \sigma_B$ uniformly at random. Then the expected Spearman's Footrule Distance ($F$) is:
\begin{align*}
\mathbb{E}[F(\sigma_A, \sigma_B)] &= \mathbb{E}[\sum^n_{i=1} |\sigma_A(i) - \sigma_B(i)|] \\
&= \sum^n_{i=1} \mathbb{E}[|\sigma_A(i) - \sigma_B(i)|] \\
&=  \sum^n_{i=1} \frac{n+1}{3} \\
&=  \frac{n(n+1)}{3} 
\end{align*}
Normalizing this by the maximum possible $F$ gives:
\begin{align*}
    \frac{\frac{n(n+1)}{3}}{\frac{n^2}{2}} &= \frac{2(n+1)}{3n} 
\end{align*}
Thus, for uncorrelated rankings of length \nmodels, $\tilde{F}\simeq 0.73$; for a perfectly correlated pair of rankings, $\tilde{F}=0$; and for perfectly anti-correlated rankings, $\tilde{F}=1$.

\subsection{Additional Details on Measuring Feedback Quality}
\label{app:fq}

\paragraph{Automatic classification of directional correctness.} We use \gpt{} to classify the feedback into two classes: (1) the feedback claimed that the solution was correct or (2) the feedback claimed that the solution was incorrect. 
As some feedback claims that the ``logic'' of the solution is correct, but then states that it is missing critical edge cases or input/output formatting, we also apply rule-based string matching to re-classify such feedback as incorrect.
We then compare the feedback to the actual \textsc{TCA} performance to classify it into \hqf{} vs. \lqf{} feedback.

\paragraph{Other metrics of feedback quality.} We considered two other metrics of feedback quality. 
First, we attempted to consider the increase in probability over either the ground-truth solution or the full question, comparing the \cm 's solution with and without feedback. 
However, we found that this measure was too noisy to impart any meaningful value. 

We also attempted to prompt \gpt{} to classify feedback relevance (to either the ground-truth solution or full question) on a scale of 1-5. 
However, this measure was also noisy, not to mention hard to define in the prompt, as even humans would struggle to distinguish between, for example, a "2" vs. a "3" in relevance.

\subsection{Performance Tables for Static vs. Interactive Settings}
\label{app:full_results}
\input{tables/full_results}
In this section, we provide tables for the performance of models across static and interactive settings, including all feedback types and baselines.
Table \ref{tables:apps-interview-full} gives the TCA of APPS (Interview), Table \ref{tables:apps-intro-full} gives the TCA of APPS (Introductory). 
Table \ref{tables:livecodebench-full} gives the TCA of LiveCodeBench and Table \ref{tables:classeval-full}.

\subsection{Additional Tables}
All additional tables -- including information about feedback quality by dataset, steerability metrics, and ranking distance metrics --- can be found in this section. 
\label{app:average_feedback_quality_per_step}

Table \ref{tables:apps-avg-feedback-quality}, Table \ref{tables:livecodebench-avg-feedback-quality}, Table \ref{tables:classeval-avg-feedback-quality} have the average directional correctness of each setting and the number of steps it takes to reach a solution with 100\% TCA. We partition the analysis by model and by feedback setting.

Table \ref{tables:steerability-edit-distance} measures the average number of edits made by each model for each feedback. 
Table \ref{tables:steerability-test-case} measures the average number of test cases flipped by each feedback setting.

Tables \ref{tables:normalized-spearmans-footrule-dist} gives the normalized Spearman's Footrule distance of each setting's ranking compared to the \vanilla{} setting.

\input{tables/average_feedback_quality_per_step}

\input{tables/feedback_vs_steerability}
\input{tables/ranking_spearmans_footrule}

%% file: tables/feedback_examples.tex
\begin{table*}
\centering
\resizebox{\textwidth}{!}{
\begin{tabular}{ll}
\toprule
\textbf{Feedback Style}       & \textbf{Generated Feedback on APPS (Interview) }                                                                                                                                                                                                                                                                                                                                                                                                                                                                                                                                                                                                                                                                                     \\ \hline
\sent & \begin{tabular}[c]{@{}l@{}}The current solution doesn't account for the cyclic nature of the strings and may count duplicates,\\so we need to implement a proper cyclic string comparison and ensure uniqueness in the count.\end{tabular}                                                                                                                                                                                                                                                                                                                                                                                                                                                                                \\\hline
\para & \begin{tabular}[c]{@{}l@{}}The current solution misunderstands the problem. It's not counting distinct cyclic strings correctly, \\ as it doesn't consider all possible binary strings of length n that contain s as a substring. \\ The solution should use a more sophisticated approach, possibly involving dynamic programming \\ or KMP algorithm, to generate and count all valid cyclic strings. It needs to handle cases where s \\ can appear at any position in the cyclic string, not just at the beginning. The current implementation\\  also doesn't account for the cyclical nature of the strings properly. A complete overhaul of the \\ algorithm is needed to correctly solve this problem.\end{tabular} \\\hline
\cf   & \begin{tabular}[c]{@{}l@{}}The main issue is that this solution doesn't correctly handle cyclic strings. The line:\\ \\ ```python\\ if s in t or s in t+t:\\ ```\\ \\ should be replaced with:\\ \\ ```python\\ if s in (t * 2):\\ ```\\ \\ This change ensures that all possible cyclic shifts are checked. Additionally, the solution\\  doesn't account for the fact that cyclic shifts of the same string should be counted as one distinct string.\\ A more fundamental restructuring of the algorithm is needed to correctly solve the problem as \\ described in the full problem statement.\end{tabular}                                                                                                            \\\hline
\ir{} & \begin{tabular}[c]{@{}l@{}} Question: Calculate the number of unique cyclical binary strings of length n that include \\ a given binary string s as a substring. A cyclical string contains s if any  rotation of the string \\ includes s. Two cyclical strings are different if they're not rotations of each other. Return the count modulo $10^9 + 7$ \end{tabular} \\ \hline
\end{tabular}
}
\caption{Example feedback from \sonnet{}, given in response to a proposed solution for APPS Interview (Question \#42)}
\label{tables:sample-feedback}
\end{table*}

%% file: figtext/4o_user_plot.tex
\begin{figure}[t]
    \includegraphics[width=0.98\columnwidth]{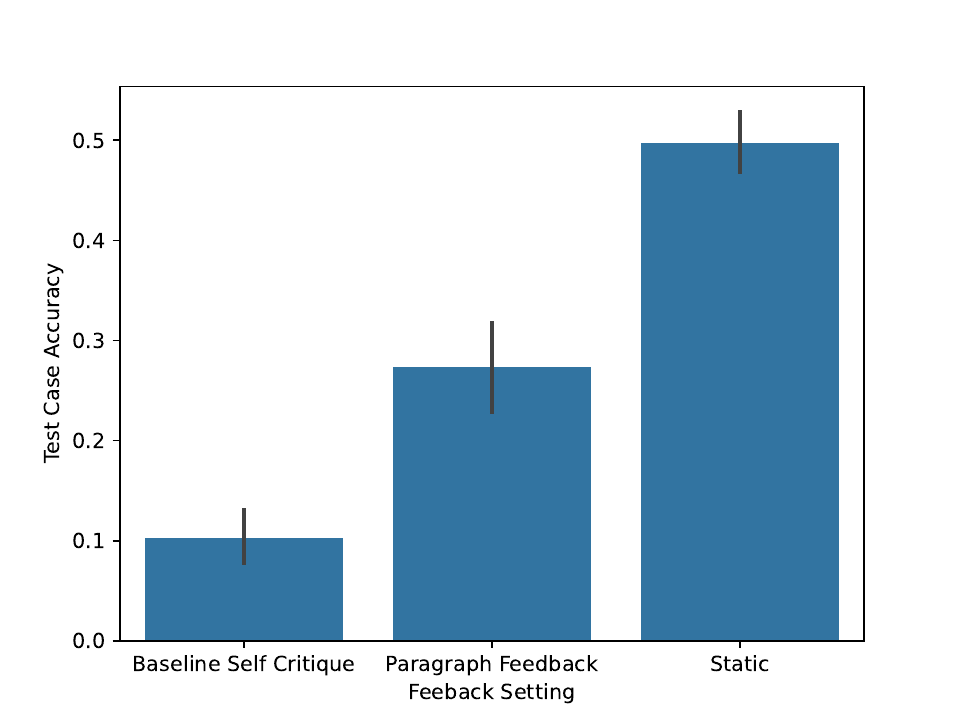}
    \caption{
        APPS (Interview) test case accuracy with \sonnet{} as the coding model and \texttt{GPT-4o-mini} as the \user{} model providing feedback. 
    }
    \vspace{-5pt}
    
    \label{fig:4o_user_w_sonnet_apps}
\end{figure}

%% file: prompts/user_system.tex
\begin{figure*}[ht]
    \centering
    \noindent\fbox{
        \parbox{\textwidth}{\vspace{0.1cm}
  You are an expert human programmer who is using a coding assistant to write code in order to solve some programming puzzles. The coding assistant has completed a potential solution to the problem, but needs your help to make adjustments to the code. \vspace{0.2cm}

  You have access to the full question, including formatting instructions and some test cases. You also have access to a natural language description of the correct solution. The coding assistant has access to a summarized, less detailed version of the problem, but only you have access to the full problem. This means that the code assistant may need additional information on how the code should work or how its output should be formatted.\vspace{0.2cm}

  Here is the description of the programming problem:
    \vspace{0.4cm}

    \bluetext{\{\textit{full question}\}}\vspace{0.4cm}
    
  Here is a description of the correct solution:
\vspace{0.4cm}

    \bluetext{\{\textit{solution info}\}}\vspace{0.4cm}
        }
        
    }
    \caption{System prompt given to \user{} model. \bluetext{Blue text} indicates that the relevant text would be inserted at that location in the prompt.}
    \label{prompts:user-sys-prompt}
\end{figure*}

%% file: prompts/user_para.tex
\begin{figure*}[ht]
    \centering
    \noindent\fbox{
        \parbox{\textwidth}{\vspace{0.1cm}

  Your goal is to provide feedback about the solution that you think would help the assistant fix or adjust the code. This feedback should be purely about the function of the code, not its aesthetics or nonessential structure (e.g. do not make a suggestion regarding optimization or other choices that would not change how the code behaves). The coding assistant will use this feedback to help generate the next version of the code. Write one paragraph of feedback that you think would be most helpful to the coding assistant. Do not write more than 100 words.\vspace{0.2cm}

  Here is the code assistant's solution.

  \vspace{0.4cm}
\textasciigrave\textasciigrave\textasciigrave python\\
  \vspace{0.2cm}
\bluetext{\textit{\{full solution\}}}\\
    \vspace{0.2cm}
  \textasciigrave\textasciigrave\textasciigrave
\vspace{0.4cm}

  Please provide one paragraph of feedback that would best help the coding assistant write a better version of the solution. Don't copy the code. Just write your reply.

        }
        
    }
    \caption{Prompt given to \user{} model to get \para{} feedback. \bluetext{Blue text} indicates that the relevant text would be inserted at that location in the prompt.}
    \label{prompts:user-para-prompt}
\end{figure*}

%% file: prompts/user_sent.tex
\begin{figure*}[ht]
    \centering
    \noindent\fbox{
        \parbox{\textwidth}{\vspace{0.1cm}

 Your goal is to provide feedback about the solution that you think would help the assistant fix or adjust the code. This feedback should be purely about the function of the code, not its aesthetics or nonessential structure (e.g. do not make a suggestion regarding optimization or other choices that would not change how the code behaves). The coding assistant will use this feedback to help generate the next version of the code. Write one sentence of feedback that you think would be most helpful to the coding assistant. Do not write more than 50 words.
  \vspace{0.2cm}

  Here is the code assistant's solution.

  \vspace{0.4cm}
\textasciigrave\textasciigrave\textasciigrave python\\
  \vspace{0.2cm}
\bluetext{\textit{\{full solution\}}}\\
    \vspace{0.2cm}
  \textasciigrave\textasciigrave\textasciigrave
\vspace{0.4cm}

Please provide one sentence of feedback that would best help the coding assistant write a better version of the solution. Don't copy the code. Just write your reply.

        }
        
    }
    \caption{Prompt given to \user{} model to get \sent{} feedback. \bluetext{Blue text} indicates that the relevant text would be inserted at that location in the prompt.}
    \label{prompts:user-sent-prompt}
\end{figure*}

%% file: prompts/user_code.tex
 \begin{figure*}[ht]
    \centering
    \noindent\fbox{
        \parbox{\textwidth}{\vspace{0.1cm}

  Your goal is to provide feedback about the solution that you think would help the assistant fix or adjust the code. This feedback should be purely about the function of the code, not its aesthetics or nonessential structure (e.g. do not make a suggestion regarding optimization or other choices that would not change how the code behaves). The coding assistant will use this feedback to help generate the next version of the code. Point out specific lines of the code that are incorrect and explain why. Do not write more than 100 words.\vspace{0.2cm}

  Here is the code assistant's solution.

  \vspace{0.4cm}
  
\textasciigrave\textasciigrave\textasciigrave python

  \vspace{0.2cm}
  
\bluetext{\textit{\{full solution\}}}

    \vspace{0.2cm}
    
  \textasciigrave\textasciigrave\textasciigrave
  
\vspace{0.4cm}

  Please point out specific lines of the code that are incorrect and give the corrected version. Make sure you copy paste the specific line in the solution which is incorrect! You should write both the original line (exactly as found in the solution) and also write what line it should be replaced with.
        }
        
    }
    \caption{Prompt given to \user{} model to get \cf. \bluetext{Blue text} indicates that the relevant text would be inserted at that location in the prompt.}
    \label{prompts:user-code-prompt}
\end{figure*}

%% file: prompts/user_input_refine.tex
 \begin{figure*}[ht]
    \centering
    \noindent\fbox{
        \parbox{\textwidth}{\vspace{0.1cm}

  Here is the code assistant's solution.
  \vspace{0.4cm}
  
\textasciigrave\textasciigrave\textasciigrave python

  \vspace{0.2cm}
  
\bluetext{\textit{\{full solution\}}}

    \vspace{0.2cm}
    
  \textasciigrave\textasciigrave\textasciigrave
  
\vspace{0.4cm}

  Here is the previous version of the question.
  \vspace{0.4cm}
  
  \bluetext{\textit{\{underspecified question\}}}
  
  \vspace{0.4cm}
  
Please rewrite the question so as to provide an updated question of similar length which would help the model generate a better version of the code. Make sure you don't make the new question longer than the older version! Begin your response with "Question:" and don't add any extra text to the end.
        }
        
    }
    \caption{Prompt given to \user{} model to get \ir{} feedback. \bluetext{Blue text} indicates that the relevant text would be inserted at that location in the prompt.}
    \label{prompts:user-input-refine-prompt}
\end{figure*}

%% file: prompts/asst_vanilla_apps_lcb.tex
 \begin{figure*}[ht]
    \centering
    \noindent\fbox{
        \parbox{\textwidth}{\vspace{0.1cm}
  You are a coding assistant who is writing code in order to solve some programming puzzles. 
  \vspace{0.2cm}
  
  You will be provided with a summary of the problem. You may also be provided with some starter code that you need to complete.
  
\vspace{0.2cm}

  Your goal is to complete the code so as to solve the problem. Do not add anything else to the code, including natural language that is not part of the code or comments. Your generation should be ready to run without needing any modifications.
  
\vspace{0.2cm}

  \orangetext{APPS: Keep in mind that for this dataset, the results should be printed to stdout, so don't just write a function without calling it or printing something to stdout.}
  
\vspace{0.2cm}

  Here is the programming problem description:
  \vspace{0.4cm}

  \bluetext{\textit{\{underspec question\}}}
  \vspace{0.4cm}
  
  Enclose your solution in a markdown block beginning with \textasciigrave\textasciigrave\textasciigrave python. When you are ready to submit your response, please end the markdown block with \textasciigrave\textasciigrave\textasciigrave on a new line.
  
  \vspace{0.4cm}

    \tantext{
 \textasciigrave\textasciigrave\textasciigrave python

  \vspace{0.2cm}
  
\textit{\{partial solution\}}
}

        }
        
    }
    \caption{Prompt given to \cm{} in the \vanilla{}, \baseline{}, and \ir{} settings for APPS and LiveCodeBench. \bluetext{Blue text} indicates that the relevant text would be inserted at that location in the prompt. \orangetext{Orange text} is the APPS-specific formatting instructions. \tantext{Red text} is prefill for the \cm{}. }
    \label{prompts:asst-vanilla_apps_lcb}
\end{figure*}

%% file: prompts/asst_nl_para_sent_apps.tex
 \begin{figure*}[ht]
    \centering
    \noindent\fbox{
        \parbox{\textwidth}{\vspace{0.1cm}
  You are a coding assistant who is writing code in order to solve some programming puzzles. 
  \vspace{0.2cm}
  
  You will be provided with a summary of the problem. You may also be provided with some starter code that you need to complete.
  
\vspace{0.2cm}

  Your goal is to complete the code so as to solve the problem. Do not add anything else to the code, including natural language that is not part of the code or comments. Your generation should be ready to run without needing any modifications.
  
\vspace{0.2cm}

  \orangetext{APPS: Keep in mind that for this dataset, the results should be printed to stdout, so don't just write a function without calling it or printing something to stdout.}
  
\vspace{0.2cm}

  Here is the programming problem description:
  \vspace{0.4cm}

  \bluetext{\textit{\{underspec question\}}}
  \vspace{0.4cm}

  Here is your last version of the code:

 \vspace{0.4cm}
 
  \textasciigrave\textasciigrave\textasciigrave python
  
   \vspace{0.2cm}
   
   \bluetext{\textit{\{prev solution\}}}
   
   \vspace{0.2cm}
   
  \textasciigrave\textasciigrave\textasciigrave
  \vspace{0.4cm}
  
  The user provided the following feedback on the code:
    \vspace{0.4cm}

  \bluetext{\textit{\{user response\}}}
  
\vspace{0.4cm}
  You can choose to use this response, or you can choose to ignore it. Only incorporate the information that you think is relevant and helpful into the code.

  Enclose your solution in a markdown block beginning with \textasciigrave\textasciigrave\textasciigrave python. When you are ready to submit your response, please end the markdown block with \textasciigrave\textasciigrave\textasciigrave on a new line.
  
  \vspace{0.4cm}

    \tantext{
 \textasciigrave\textasciigrave\textasciigrave python

  \vspace{0.2cm}
  
}

        }
        
    }
    \caption{Prompt given to \cm{} in the \para{}, \sent{}, \cf{} setting for APPS and LiveCodeBench. \bluetext{Blue text} indicates that the relevant text would be inserted at that location in the prompt. \orangetext{Orange text} is the APPS specific formatting instructions. \tantext{Red text} is prefill for the \cm{}. }
    \label{prompts:asst-nl_para_sent_apps_lcb}
\end{figure*}

%% file: prompts/asst_vanilla_classeval.tex
\begin{figure*}[ht]
    \centering
    \noindent\fbox{
        \parbox{\textwidth}{\vspace{0.1cm}
  You are a coding assistant who is writing code in order to fill in the skeleton of a python class. 
  
    \vspace{0.2cm}
    
  You will be provided with the skeleton of the class with function names and docstrings. You may also be provided with some functions already completed.

\vspace{0.2cm}

  Your goal is to complete the code according to the docstrings. Do not add anything else to the code, including natural language that is not part of the code or comments. Your generation should be ready to run without needing any modifications. 

  \vspace{0.2cm}

  Here is the skeleton:

  \vspace{0.4cm}

  \bluetext{\textit{\{underspec question\}}}
  
  \vspace{0.4cm}
  
  Enclose your solution in a markdown block beginning with \textasciigrave\textasciigrave\textasciigrave python. When you are ready to submit your response, please end the markdown block with \textasciigrave\textasciigrave\textasciigrave on a new line.
  
  \vspace{0.4cm}

    \tantext{
 \textasciigrave\textasciigrave\textasciigrave python

  \vspace{0.2cm}
  
\textit{\{partial solution\}}
}

        }
        
    }
    \caption{Prompt given to \cm{} in the \vanilla{}, \baseline{}, and \ir{} settings for ClassEval. \bluetext{Blue text} indicates that the relevant text would be inserted at that location in the prompt. \tantext{Red text} is prefill for the \cm{}. }
    \label{prompts:asst-classeval_vanilla}
\end{figure*}

%% file: prompts/asst_nl_para_sent_classeval.tex
 \begin{figure*}[ht]
    \centering
    \noindent\fbox{
        \parbox{\textwidth}{\vspace{0.1cm}
  You are a coding assistant who is writing code to fill in some incomplete classes. 
  \vspace{0.2cm}
  
    You will be provided with the skeleton of the class with function names and docstrings. You may also be provided with some functions already completed.
  
\vspace{0.2cm}

    Your goal is to complete the code according to the docstrings. Do not add anything else to the code, including natural language that is not part of the code or comments. Your generation should be ready to run without needing any modifications. 
  
\vspace{0.2cm}

  Here is the skeleton:
  \vspace{0.4cm}

  \bluetext{\textit{\{underspec question\}}}
  \vspace{0.4cm}

  Here is your last version of the skeleton:

 \vspace{0.4cm}
 
  \textasciigrave\textasciigrave\textasciigrave python
  
   \vspace{0.2cm}
   
   \bluetext{\textit{\{prev solution\}}}
   
   \vspace{0.2cm}
   
  \textasciigrave\textasciigrave\textasciigrave
  \vspace{0.4cm}
  The user provided the following feedback on the code:
    \vspace{0.4cm}

  \bluetext{\textit{\{user response\}}}
\vspace{0.4cm}
  You can choose to use this response, or you can choose to ignore it. Only incorporate the information that you think is relevant and helpful into the code.

  Enclose your solution in a markdown block beginning with \textasciigrave\textasciigrave\textasciigrave python. When you are ready to submit your response, please end the markdown block with \textasciigrave\textasciigrave\textasciigrave on a new line.
  
  \vspace{0.4cm}

    \tantext{
 \textasciigrave\textasciigrave\textasciigrave python

  \vspace{0.2cm}
  
}

        }
        
    }
    \caption{Prompt given to \cm{} in the \para{}, \sent{}, \cf{} setting for Classeval. \bluetext{Blue text} indicates that the relevant text would be inserted at that location in the prompt. \tantext{Red text} is prefill for the \cm{}. }
    \label{prompts:asst-nl_para_sent_classeval}
\end{figure*}

%% file: prompts/summarization_apps_lcb.tex
 \begin{figure*}[ht]
    \centering
    \noindent\fbox{
        \parbox{\textwidth}{ \vspace{0.1cm}
        Summarize the question with header 'FORMAL QUESTION' using only natural language. Write your summary under 'SUMMARY'
        
  Do not use any variable or function names from the question. Do not write any code.

   "You are given a coding/algorithmic question below. Your goal is to come up with a \bluetext{\textit{\{sent length\}}} sentence summary using natural language to describe the problem.
   
  "Here are examples of a question and a summary labeled 'EX QUESTION' and 'EX SUMMARY'. 
  Format your summary of 'FORMAL QUESTION' in a similar way to these summaries using \bluetext{\textit{\{sent length\}}} sentence(s).

    \vspace{0.4cm}

  \#\#\#EX QUESTION
  
    \vspace{0.2cm}
    
  Example question 1

    \vspace{0.4cm}

\#\#\#EX SUMMARY

  \vspace{0.2cm}
  
 Example summary 1

  \vspace{0.4cm}

  \#\#\#EX QUESTION

  \vspace{0.2cm}
  
  Example question 2

  \vspace{0.4cm}

\#\#\#EX SUMMARY

  \vspace{0.2cm}

Example summary 2

  \vspace{0.4cm}
  
  \#\#\#EX QUESTION

  \vspace{0.2cm}
  
  Example question 3

  \vspace{0.4cm}

\#\#\#EX SUMMARY

  \vspace{0.2cm}

 Example summary 3

$\vdots$

  \vspace{0.4cm}

\#\#\#EX QUESTION

  \vspace{0.2cm}
  
 Example question 11

  \vspace{0.4cm}

  \#\#\#EX SUMMARY

  \vspace{0.2cm}
  
  Example summary 11

  \vspace{0.4cm}

  \#\#\#FORMAL QUESTION
  
    \vspace{0.2cm}
    
  \bluetext{\textit{\{question\}}}
  
  \vspace{0.4cm}
  
  \#\#\#SUMMARY
  
    \vspace{0.2cm}
    
  <YOUR SUMMARY HERE>
 
        }
        
    }
    \caption{Format for the 11 shot prompt we use to generate summaries in APPS and LiveCodeBench problems. \bluetext{Blue text} indicates that the relevant text would be inserted at that location in the prompt.}
    \label{prompts:summarization_apps_lcb}
\end{figure*}

%% file: prompts/summarization_classeval.tex
 \begin{figure*}[ht]
    \centering
    \noindent\fbox{
        \parbox{\textwidth}{\vspace{0.1cm}
  Do not reference any variables or function names. Do not write any code or examples of behavior.

    \vspace{0.4cm}

  You are given a method signature and a docstring below. Write a short, one sentence summary of the docstring. Do not use code or examples in your summary. Retain only the key information. Do not use more than 15 words.

    \vspace{0.4cm}

  \#\# SIGNATURE and DOCSTRING
    \bluetext{\textit{\{function\}}}

  \#\# SUMMARY
  <YOUR SUMMARY HERE>

        }
        
    }
    \caption{Prompt we use to generate summarized docstring for each function in ClassEval skeletons. \bluetext{Blue text} indicates that the relevant text would be inserted at that location in the prompt.}
    \label{prompts:summarization_classeval}
\end{figure*}

%% file: tables/full_results.tex
\begin{table*}
\centering
\resizebox{\textwidth}{!}{
\begin{tabular}{ccccccc}
\toprule 
\textbf{Model}  & \textbf{\vanilla} & \textbf{\baseline} & \textbf{\sent} & \textbf{\para} & \textbf{\cf} & \textbf{\ir} \\ \hline
\gpt & 0.498 (0.016) & 0.068 (0.007) & 0.422 (0.016) & 0.544 (0.016) & \textbf{0.598 (0.016) }& 0.488 (0.016)\\
\aya &\textbf{ 0.261 (0.014)} & 0.009 (0.002) & 0.131 (0.011) & 0.259 (0.015) & 0.235 (0.014) & 0.214 (0.013)\\
\deepseek & \textbf{0.616 (0.015)} & 0.048 (0.007) & 0.449 (0.016) & 0.512 (0.023) & 0.521 (0.026) & 0.442 (0.019)\\
\gemmaLarge & 0.409 (0.013) & 0.007 (0.002) & 0.351 (0.016) & \textbf{0.556 (0.016)} & 0.51 (0.017) & 0.403 (0.015)\\
\gemmaSmall & 0.177 (0.01) & 0.009 (0.002) & 0.039 (0.006) & 0.084 (0.009) &\textbf{ 0.299 (0.016)} & 0.029 (0.004)\\
\llama & 0.253 (0.012) & 0.025 (0.003) & 0.236 (0.014) & 0.402 (0.015) & \textbf{0.453 (0.016)} & 0.215 (0.012)\\
\qwenSmall & 0.369 (0.014) & 0.026 (0.004) & 0.283 (0.014) & 0.423 (0.016) & \textbf{0.495 (0.016)} & 0.336 (0.015)\\
\qwenLarge & 0.542 (0.015) & 0.039 (0.004) & 0.5 (0.016) & 0.582 (0.015) & \textbf{0.605 (0.015)} & 0.503 (0.015)\\
\reka & 0.164 (0.011) & 0.018 (0.003) & 0.224 (0.013) & 0.303 (0.015) & \textbf{0.4 (0.016)} & 0.157 (0.011)\\
\sonnet & 0.59 (0.014) & 0.114 (0.009) & 0.571 (0.015) & \textbf{0.654 (0.014)} & 0.627 (0.015) & 0.62 (0.014)\\   
\bottomrule
\end{tabular}
}
\caption{Average TCA of each model with standard error on APPS Interview questions. 
}
\label{tables:apps-interview-full}
\end{table*}

\begin{table*}
\centering
\resizebox{\textwidth}{!}{
\begin{tabular}{ccccccc}
\toprule 
\textbf{Model}  & \textbf{\vanilla} & \textbf{\baseline} & \textbf{\sent} & \textbf{\para} & \textbf{\cf} & \textbf{\ir} \\ \hline
\gpt & 0.412 (0.015) & 0.071 (0.007) & 0.392 (0.016) & 0.512 (0.016) & \textbf{0.539 (0.016)} & 0.409 (0.016)\\
\aya & 0.182 (0.008) & 0.004 (0.001) & 0.069 (0.006) & \textbf{0.179 (0.009)} & 0.177 (0.009) & 0.142 (0.007)\\
\deepseek{} & - & - & - & - & - & - \\
\gemmaLarge & 0.322 (0.012) & 0.018 (0.003) & 0.228 (0.013) & \textbf{0.454 (0.016)} & 0.41 (0.016) & 0.299 (0.013)\\
\gemmaSmall & 0.131 (0.008) & 0.013 (0.003) & 0.023 (0.004) & 0.045 (0.006) & \textbf{0.213 (0.014)} & 0.014 (0.003)\\
\llama & 0.205 (0.01) & 0.028 (0.004) & 0.12 (0.009) & 0.292 (0.014) & \textbf{0.402 (0.015)} & 0.153 (0.01)\\
\qwenSmall & 0.28 (0.012) & 0.034 (0.004) & 0.186 (0.012) & 0.27 (0.014) & \textbf{0.392 (0.016)} & 0.2 (0.011)\\
\qwenLarge & 0.348 (0.024) & 0.038 (0.007) & 0.376 (0.023) & 0.479 (0.025) & \textbf{0.486 (0.025) }& 0.37 (0.023)\\
\reka & 0.124 (0.009) & 0.018 (0.003) & 0.141 (0.01) & 0.275 (0.014) & \textbf{0.34 (0.015)} & 0.11 (0.009)\\
\sonnet & 0.497 (0.014) & 0.073 (0.007) & 0.468 (0.015) & \textbf{0.556 (0.015)} & 0.53 (0.015) & 0.492 (0.015)\\
\bottomrule
\end{tabular}
}
\caption{Average TCA of each model with standard error in each setting for APPS introductory. 
\deepseek{} is missing for this setting due to rate limits on the API that impeded evaluation. }
\label{tables:apps-intro-full}
\end{table*}

\begin{table*}
\centering
\resizebox{\textwidth}{!}{
\begin{tabular}{ccccccc}
\toprule 
\textbf{Model}  & \textbf{\vanilla} & \textbf{\baseline }& \textbf{\sent} & \textbf{\para }& \textbf{\cf} & \textbf{\ir} \\ \hline
\gpt & \textbf{0.8 (0.023)} & 0.323 (0.026) & 0.767 (0.024) & 0.745 (0.024) & 0.702 (0.026) & 0.23 (0.025)\\
\aya & 0.522 (0.027) & 0.235 (0.022) & 0.482 (0.027) & \textbf{0.632 (0.026)} & 0.483 (0.028) & 0.134 (0.02)\\
\deepseek & \textbf{0.944 (0.013)} & 0.332 (0.026) & 0.841 (0.02) & 0.849 (0.019) & 0.756 (0.024) & 0.345 (0.028)\\
\gemmaLarge & \textbf{0.766 (0.021)} & 0.282 (0.025) & 0.67 (0.026) & \textbf{0.766 (0.023)} & 0.62 (0.026) & 0.119 (0.019)\\
\gemmaSmall & 0.347 (0.023) & 0.173 (0.019) & 0.196 (0.021) & 0.285 (0.024) & \textbf{0.524 (0.027)} & 0.028 (0.008)\\
\llama & 0.587 (0.025) & 0.237 (0.022) & 0.538 (0.027) & 0.588 (0.026) & \textbf{0.647 (0.026) }& 0.203 (0.023)\\
\qwenSmall & \textbf{0.755 (0.021)} & 0.27 (0.024) & 0.621 (0.027) & 0.678 (0.026) & 0.629 (0.027) & 0.189 (0.023)\\
\qwenLarge &\textbf{ 0.961 (0.007)} & 0.309 (0.025) & 0.809 (0.021) & 0.747 (0.024) & 0.712 (0.025) & 0.237 (0.025)\\
\reka & 0.617 (0.025) & 0.246 (0.023) & 0.583 (0.026) & \textbf{0.636 (0.026)} & 0.629 (0.027) & 0.111 (0.018)\\
\sonnet & \textbf{0.937 (0.012)} & 0.387 (0.026) & 0.832 (0.02) & 0.821 (0.02) & 0.735 (0.025) & 0.395 (0.029)\\   
\bottomrule
\end{tabular}
}
\caption{Average TCA of each model with standard error on a subset LiveCodeBench questions. To provide code solutions to the \user{} model, we select questions which \sonnet{} solves perfectly within two attempts, using the generated solution as ground truth. 
}
\label{tables:livecodebench-full}
\end{table*}

\begin{table*}
\centering
\resizebox{\textwidth}{!}{
\begin{tabular}{ccccccc}
\toprule 
\textbf{Model}  & \textbf{\vanilla} & \textbf{\baseline} & \textbf{\sent} & \textbf{\para} & \textbf{\cf} & \textbf{\ir} \\ \hline
\gpt & 0.839 (0.005) & 0.561 (0.018) & 0.836 (0.014) & 0.848 (0.014) & \textbf{0.895 (0.011)} & 0.789 (0.014) \\
\aya & 0.718 (0.007)  & 0.305 (0.019) & 0.596 (0.021)  & 0.597 (0.022) &\textbf{0.781 (0.018)} & 0.675 (0.016)  \\
\deepseek & 0.849 (0.005) & 0.559 (0.018) & 0.837 (0.013) & 0.867 (0.012) & \textbf{0.889 (0.012)} & 0.806 (0.013) \\
\gemmaLarge & 0.704 (0.008) & 0.563 (0.018) & 0.776 (0.015) & 0.815 (0.015) & \textbf{0.867 (0.014)} & 0.735 (0.015)  \\
\gemmaSmall & 0.350 (0.008) & 0.303 (0.017) & 0.375 (0.018) & 0.390 (0.018) &\textbf{ 0.687 (0.019) } & 0.340 (0.017)  \\
\llama & 0.636 (0.007) & 0.443 (0.019) & 0.653 (0.019) & 0.710 (0.019) &\textbf{ 0.773 (0.017) }& 0.701 (0.018)  \\
\qwenSmall & 0.714 (0.006) & 0.502 (0.019) & 0.605 (0.018) & 0.757 (0.017) & \textbf{0.819 (0.014)} & 0.697 (0.016) \\
\qwenLarge & 0.816 (0.006) & 0.542 (0.018) & 0.815 (0.014) & 0.832 (0.015) & \textbf{0.876 (0.012)} & 0.778 (0.014) \\
\reka & 0.670 (0.007) & 0.471 (0.027) & 0.096 (0.015) & 0.109 (0.016) & 0.112 (0.016) & \textbf{0.600 (0.019)} \\
\sonnet & 0.833 (0.006) & 0.564 (0.019) & 0.821 (0.014) & 0.865 (0.013) & \textbf{0.881 (0.013)}  & 0.803 (0.014) \\

\bottomrule
\end{tabular}
}
\caption{Average TCA with standard error in each setting for ClassEval.}
\label{tables:classeval-full}
\end{table*}

%% file: tables/average_feedback_quality_per_step.tex
\begin{table*}
\centering
\resizebox{\textwidth}{!}{
\begin{tabular}{cccccccc}
\toprule 
\textbf{Model} & \textbf{Feedback Type} & \textbf{Average Steps to Correct Solution} & \textbf{Average Directional Correctness} \\
\hline
 & Code Feedback & 2.981 & 0.937  \\
\gpt & Paragraph & 3.003 & 0.896  \\
 & Sentence & 3.436 & 0.850  \\
\hline
 & Code Feedback & 3.586 & 0.928  \\
\aya & Paragraph & 3.687 & 0.926  \\
 & Sentence & 3.886 & 0.901  \\
\hline
 & Code Feedback & 3.207 & 0.928  \\
\deepseek & Paragraph & 3.132 & 0.886  \\
 & Sentence & 3.484 & 0.873  \\
\hline
 & Code Feedback & 3.296 & 0.948  \\
\gemmaLarge & Paragraph & 3.259 & 0.920  \\
 & Sentence & 3.630 & 0.895  \\
\hline
 & Code Feedback & 3.661 & 0.978  \\
\gemmaSmall & Paragraph & 3.947 & 0.982  \\
 & Sentence & 3.977 & 0.950  \\
\hline
 & Code Feedback & 3.349 & 0.963  \\
\llama & Paragraph & 3.567 & 0.923  \\
 & Sentence & 3.850 & 0.897  \\
\hline
 & Code Feedback & 3.328 & 0.948  \\
\qwenSmall & Paragraph & 3.453 & 0.909  \\
 & Sentence & 3.724 & 0.899  \\
\hline
 & Code Feedback & 3.082 & 0.952  \\
\qwenLarge & Paragraph & 3.058 & 0.900  \\
 & Sentence & 3.411 & 0.876  \\
\hline
 & Code Feedback & 3.520 & 0.935  \\
\reka & Paragraph & 3.634 & 0.927  \\
 & Sentence & 3.854 & 0.879  \\
\hline
 & Code Feedback & 3.035 & 0.914  \\
\sonnet & Paragraph & 2.882 & 0.865  \\
 & Sentence & 3.299 & 0.834  \\
\bottomrule
\end{tabular}
}
\caption{Average directional correctness of feedback and the average number of steps required to reach 100\% TCA on the APPS dataset.}
\label{tables:apps-avg-feedback-quality}
\end{table*}

\begin{table*}
\centering
\resizebox{\textwidth}{!}{
\begin{tabular}{cccccccc}
\toprule 
\textbf{Model} & \textbf{Feedback Type} & \textbf{Average Steps to Correct Solution} & \textbf{Average Directional Correctness} \\
\hline
 & Code Feedback & 2.697 & 0.736  \\
\gpt & Paragraph & 2.337 & 0.902  \\
 & Sentence & 2.416 & 0.821  \\
\hline
 & Code Feedback & 2.856 & 0.852  \\
\aya & Paragraph & 2.553 & 0.945  \\
 & Sentence & 3.097 & 0.898  \\
\hline
 & Code Feedback & 2.487 & 0.785  \\
\deepseek & Paragraph & 1.808 & 0.891  \\
 & Sentence & 2.063 & 0.806  \\
\hline
 & Code Feedback & 3.089 & 0.809  \\
\gemmaLarge & Paragraph & 2.246 & 0.912  \\
 & Sentence & 2.685 & 0.896  \\
\hline
 & Code Feedback & 3.229 & 0.927  \\
\gemmaSmall & Paragraph & 3.640 & 0.986  \\
 & Sentence & 3.818 & 0.963  \\
\hline
 & Code Feedback & 2.773 & 0.888  \\
\llama & Paragraph & 2.672 & 0.957  \\
 & Sentence & 3.169 & 0.890  \\
\hline
 & Code Feedback & 2.876 & 0.786  \\
\qwenSmall & Paragraph & 2.314 & 0.958  \\
 & Sentence & 2.847 & 0.909  \\
\hline
 & Code Feedback & 2.811 & 0.771  \\
\qwenLarge & Paragraph & 1.949 & 0.871  \\
 & Sentence & 2.219 & 0.817  \\
\hline
 & Code Feedback & 2.801 & 0.668  \\
\reka & Paragraph & 2.575 & 0.883  \\
 & Sentence & 2.871 & 0.807  \\
\hline
 & Code Feedback & 2.663 & 0.701  \\
\sonnet & Paragraph & 2.017 & 0.885  \\
 & Sentence & 2.247 & 0.859  \\
\bottomrule
\end{tabular}
}
\caption{Average directional correctness of feedback and the average number of steps required to reach 100\% TCA on the LiveCodeBench dataset.}
\label{tables:livecodebench-avg-feedback-quality}
\end{table*}

\begin{table*}
\centering
\resizebox{\textwidth}{!}{
\begin{tabular}{cccccccc}
\toprule 
\textbf{Model} & \textbf{Feedback Type} & \textbf{Average Steps to Correct Solution} & \textbf{Average Directional Correctness} \\
\hline

 & Code Feedback & 3.436 & 0.743  \\
\gpt & Paragraph & 3.546 & 0.963  \\
 & Sentence & 3.153 & 0.899  \\
\hline
 & Code Feedback & 3.977 & 0.749  \\
\aya & Paragraph & 3.987 & 0.953  \\
 & Sentence & 3.840 & 0.898  \\
\hline
 & Code Feedback & 4.000 & 0.698  \\
\deepseek & Paragraph & 4.000 & 0.969  \\
 & Sentence & 4.000 & 0.910  \\
\hline
 & Code Feedback & 3.076 & 0.789  \\
\gemmaLarge & Paragraph & 3.334 & 0.977  \\
 & Sentence & 3.661 & 0.917  \\
\hline
 & Code Feedback & 3.880 & 0.902  \\
\gemmaSmall & Paragraph & 3.874 & 0.979  \\
 & Sentence & 3.940 & 0.944  \\
\hline
 & Code Feedback & 3.389 & 0.872  \\
\llama & Paragraph & 3.806 & 0.959  \\
 & Sentence & 3.699 & 0.899  \\
\hline
 & Code Feedback & 3.684 & 0.782  \\
\qwenSmall & Paragraph & 3.724 & 0.967  \\
 & Sentence & 3.322 & 0.929  \\
\hline
 & Code Feedback & 3.536 & 0.744  \\
\qwenLarge & Paragraph & 4.000 & 0.977  \\
 & Sentence & 3.585 & 0.889  \\
\hline
 & Code Feedback & 3.978 & 0.769  \\
\reka & Paragraph & 4.000 & 0.942  \\
 & Sentence & 4.000 & 0.901  \\
\hline
 & Code Feedback & 3.664 & 0.659  \\
\sonnet & Paragraph & 3.675 & 0.947  \\
 & Sentence & 4.000 & 0.913  \\
\bottomrule
\end{tabular}
}
\caption{Average directional correctness of feedback and the average number of steps required to reach 100\% TCA on the ClassEval dataset}
\label{tables:classeval-avg-feedback-quality}
\end{table*}

\begin{table*}
\centering
\resizebox{\textwidth}{!}{
\begin{tabular}{cccccccc}
\toprule 
\textbf{Model} & \textbf{Feedback Type} & \textbf{Average Steps to Correct Solution} & \textbf{Average Directional Correctness} \\
\hline

 & Code Feedback & 2.925 & 0.904  \\
\gpt & Paragraph & 2.916 & 0.906  \\
 & Sentence & 3.293 & 0.858  \\
\hline
 & Code Feedback & 3.400 & 0.918  \\
\aya & Paragraph & 3.516 & 0.934  \\
 & Sentence & 3.738 & 0.905  \\
\hline
 & Code Feedback & 2.814 & 0.831  \\
\deepseek & Paragraph & 2.721 & 0.929  \\
 & Sentence & 3.171 & 0.886  \\
\hline
 & Code Feedback & 3.218 & 0.915  \\
\gemmaLarge & Paragraph & 3.120 & 0.925  \\
 & Sentence & 3.479 & 0.903  \\
\hline
 & Code Feedback & 3.545 & 0.969  \\
\gemmaSmall & Paragraph & 3.888 & 0.982  \\
 & Sentence & 3.936 & 0.952  \\
\hline
 & Code Feedback & 3.250 & 0.945  \\
\llama & Paragraph & 3.427 & 0.930  \\
 & Sentence & 3.708 & 0.900  \\
\hline
 & Code Feedback & 3.237 & 0.921  \\
\qwenSmall & Paragraph & 3.287 & 0.917  \\
 & Sentence & 3.663 & 0.874  \\
\hline
 & Code Feedback & 3.002 & 0.919  \\
\qwenLarge & Paragraph & 2.931 & 0.908  \\
 & Sentence & 3.265 & 0.883  \\
\hline
 & Code Feedback & 3.508 & 0.889  \\
\reka & Paragraph & 3.560 & 0.929  \\
 & Sentence & 3.761 & 0.878  \\
\hline
 & Code Feedback & 2.983 & 0.877  \\
\sonnet & Paragraph & 2.803 & 0.880  \\
 & Sentence & 3.193 & 0.849  \\
\bottomrule
\end{tabular}
}
\caption{Average directional correctness of feedback and the average number of steps required to reach 100\% TCA across all datasets.}
\label{tables:all-avg-feedback-quality}
\end{table*}

%% file: tables/feedback_vs_steerability.tex
\begin{table*}
\centering
\resizebox{\textwidth}{!}{
\begin{tabular}{cccccccc}
\toprule 
\textbf{Model} & \textbf{\sent} & \textbf{\para} & \textbf{\cf} & \textbf{\ir} \\ \hline
\gpt & 265.029 & 308.837 & 451.749 & 325.441 \\
\aya & 333.695 & 274.630 & 503.740 & 351.736 \\
\deepseek & 143.413 & 190.044 & 301.157 & 264.532 \\
\gemmaLarge & 116.412 & 108.401 & 182.592 & 108.516 \\
\gemmaSmall & 286.987 & 211.680 & 299.998 & 261.134 \\
\llama & 361.366 & 400.246 & 511.731 & 385.156 \\
\qwenSmall & 214.421 & 178.882 & 363.474 & 211.171 \\
\qwenLarge & 270.493 & 230.114 & 496.959 & 328.425 \\
\reka & 541.689 & 292.388 & 768.962 & 638.245 \\
\sonnet & 155.923 & 204.138 & 320.164 & 215.693 \\
\bottomrule
\end{tabular}
}
\caption{Surface-level steerability (as measured by edit distance) vs. feedback type across each model.}
\label{tables:steerability-edit-distance}
\end{table*}

\begin{table*}
\centering
\resizebox{\textwidth}{!}{
\begin{tabular}{cccccccc}
\toprule 
\textbf{Model} & \textbf{\sent} & \textbf{\para} & \textbf{\cf} & \textbf{\ir} \\ \hline
\gpt & 0.225 & 0.164 & 0.240 & 0.152 \\
\aya & 0.169 & 0.090 & 0.208 & 0.108 \\
\deepseek & 0.159 & 0.138 & 0.208 & 0.161 \\
\gemmaLarge & 0.149 & 0.113 & 0.199 & 0.095 \\
\gemmaSmall & 0.094 & 0.007 & 0.040 & 0.015 \\
\llama & 0.2 & 0.102 & 0.201 & 0.107 \\
\qwenSmall & 0.157 & 0.095 & 0.184 & 0.101 \\
\qwenLarge & 0.228 & 0.142 & 0.263 & 0.192 \\
\reka & 0.154 & 0.059 & 0.168 & 0.095 \\
\sonnet & 0.239 & 0.225 & 0.311 & 0.212 \\
\bottomrule
\end{tabular}
}
\caption{Behavioral-level steerability (as measured by number of test cases changed from correct to incorrect or vice-versa) vs. feedback type across each model.}
\label{tables:steerability-test-case}
\end{table*}

%% file: tables/ranking_spearmans_footrule.tex
\begin{table*}
\centering
\begin{tabular}{ccc}
\toprule 
\textbf{Dataset} & \textbf{Feedback Type} & \textbf{Normalized Spearman's Footrule Distance} \\ \hline
\multirow{4}{*}{Apps}  & Code Feedback & 0.267  \\
 & Input Refinement & 0.222  \\
 & Paragraph & 0.222  \\
 & Sentence & 0.178  \\
\hline
\multirow{4}{*}{ClassEval}  & Code Feedback & 0.222 \\
 & Input Refinement & 0.267  \\
 & Paragraph & 0.267  \\
 & Sentence & 0.267  \\
\hline
\multirow{4}{*}{LiveCodeBench} & Code Feedback & 0.356  \\
 & Input Refinement & 0.356  \\
 & Paragraph & 0.222  \\
 & Sentence & 0.044 \\

\bottomrule
\end{tabular}
\caption{Normalized Spearman's Footrule Distance when comparing each feedback setting's ranking order to the ranking order on static benchmark.}
\label{tables:normalized-spearmans-footrule-dist}
\end{table*}